\documentclass[aip,chaos,reprint,twocolumn]{revtex4-1}
\usepackage{amsmath}
\usepackage{graphicx}
\graphicspath{{./figures/}}
\usepackage{epsfig}
\usepackage{float}
\usepackage[usenames, dvipsnames]{xcolor}
\usepackage{stackengine}
\newfont{\logo}{logo10}

\newcommand{\bea}{\begin{eqnarray}}
\newcommand{\eea}{\end{eqnarray}}
\newcommand{\bes}{\begin{subequations}}
\newcommand{\ees}{\end{subequations}}

\newcommand{\sn}{{\rm sn}}

\newcommand{\sine}{{\rm sine}}
\begin{document}
\bibliographystyle{revtex4-1}
\title{Cubic-quintic nonlinear Helmholtz equation: Modulational instability, chirped elliptic and solitary waves}
\author{K. Tamilselvan}
\author{ T. Kanna}\email[Corresponding author: ]{kanna\_phy@bhc.edu.in}
\affiliation{Nonlinear Waves Research Lab, PG and Research Department of Physics, Bishop Heber College, Tiruchirappalli - 620 017, Tamil Nadu, India}
\author{A. Govindarajan}
\affiliation{Centre for Nonlinear Dynamics, School of Physics, Bharathidasan University, Tiruchirappalli - 620 024, Tamil Nadu, India}
\begin{abstract}	
We study the formation and propagation of chirped elliptic and solitary waves in cubic-quintic nonlinear Helmholtz (CQNLH) equation. This system  describes nonparaxial pulse propagation in a planar waveguide with Kerr-like and  quintic nonlinearities along with spatial dispersion originating from the nonparaxial effect that becomes dominant when the conventional  slowly varying envelope approximation (SVEA) fails. We first carry out the modulational instability (MI) analysis of a plane wave in this system by employing the linear stability analysis and investigate the influence of different physical parameters on the MI gain spectra. In particular, we show the nonparaxial parameter suppresses the conventional MI gain spectrum and also leads to a nontrivial monotonic increase in the gain spectrum near the tails of the conventional MI band, a qualitatively distinct behaviour from the standard nonlinear Schr\"odinger (NLS) system. We then study the MI dynamics by direct numerical simulations which demonstrate production of ultra-short nonparaxial pulse trains with internal oscillations and slight distortions at the wings. Following the MI dynamics, we obtain exact elliptic and solitary wave solutions using the integration method by considering physically interesting chirped traveling wave ansatz. In particular, we show that the system features intriguing chirped anti-dark, bright, gray and dark solitary waves depending upon the nature of nonlinearities. We also show that the chirping is inversely  proportional to the intensity of the optical wave. Especially, the bright and dark solitary waves exhibit unusual chirping behaviour which will have applications in nonlinear pulse compression process.
\end{abstract}
\maketitle
{\bf Nonlinear Helmholtz (NLH) equation can play a significant role in modeling a progressive miniaturization of photonic devices and plasmonics. By obeying the intrinsic higher dimensions of spatial symmetry of uniform planar waveguides, many experimental configurations could be modeled that are otherwise inaccessible in the standard SVEA. In the present work, we consider the NLH equation with cubic-quintic (CQ) nonlinearities. The implication of non-Kerr nonlinearity in the NLH system leads to a more stabilized non-paraxial pulse propagation and it can easily be realized in available materials including semiconductors and doped fibers. We study the phenomenon of MI in  the CQNLH system, where we address some peculiar features of nonparaxiality. Also, we numerically demonstrate the generation of train of ultra-short nonparaxial solitary pulses. In addition to different types of Helmholtz solitons, we present the first ever study of formation and  propagation of chirped elliptic and solitary waves, which are not reported for any NLH system before. In particular, we obtain various interesting chirped solitary profiles, which include bright, dark, gray and anti-dark solitary waves depending upon the nature of nonlinearities.  Among them, we find  chirped profiles of bright and dark solitary waves exhibit some unusual nonlinear dynamics of compression and amplification. All the analytical solutions well corroborate with direct numerical simulations.}
\section{Introduction}
Ever since the first observation of optical solitons \cite{molle} by the pioneering work of Mollenauer \emph{et al.}, studies pertaining to different roles and dynamics of optical solitons remain to be a frontier research topic in the context of nonlinear optics \cite{kivshar}. Note that such experimental vindications of optical solitons have been realized in various optical media, which include optical fibers \cite{molle2}, planar-waveguides \cite{fried}, photo-refractive media \cite{duree}, optical cavities \cite{leo},  photonic crystal fibers \cite{jin} and so on. In a nonlinear Kerr medium, such stable solitons form as a result of delicate interplay between the linear dispersion/diffraction and nonlinearity. Here,  the Kerr nonlinearity induces an intensity dependent nonlinear phase shift, which is generally referred as the self-phase modulation (SPM) effect. From a theoretical perspective, the propagation of pico-second light pulses in optical waveguides can be well described by the ubiquitous integrable nonlinear Schr\"odinger (NLS) equation, which was originally derived  by Hasegawa  and Tappert  by considering the SVEA \cite{Hasegawa0}. This NLS system features different types of nonlinear waves such as  bright solitons \cite{zakharov}, dark solitons \cite{zakharov1}, breather solutions \cite{Akhmediev0,tajiri}, Peregrine solitons \cite{kibler} and higher order rogue waves \cite{kunznetsov} depending upon the nature of nonlinearities.

Subsequently, various generalizations of the NLS system become more important when one encounters higher order nonlinear effects for ultra-short and intense pulses in real fibers. One such higher order nonlinear model is the CQNLS system and this system has received great attention due to its  experimental realization in different nonlinear materials like $AlGaAs$ semiconductor \cite{tanev}, $CdS_{x} Se_{1-x}$ doped glass \cite{acioli}, chalcogenide glasses in the ternary system $Ag$-$As$-$Se$ \cite{lawrence} and bulk single-crystal polydiacetylene para-toluene sulfonate \cite{ogusu}. Such CQ nonlinearities can be achieved by taking into account of the third and fifth order nonlinear dielectric susceptibilities $\chi^{(3)}$ and $\chi^{(5)}$, respectively. Initially, the bright solitary wave has  been studied for the coupled CQNLS system \cite{pushkarov} followed by a numerical investigation on the interaction of solitary waves along with a stability analysis \cite{cowan}. Later on, existence of bistable behaviour (undistorted pulses with the same duration but different peak power) of the bright and dark solitary waves has been studied in the CQ nonlinear media \cite{tanev1}.

All these preceding works have considered the CQNLS system derived under the SVEA (also known as paraxial approximation) in which the pulse (beam) width is broader than the carrier wavelength along with sufficiently low intensity and also the pulse (beam) is assumed to propagate along (or at near-negligible angles with) the reference axis \cite{chamorro}. If the pulse (beam) fails to satisfy at least any one of the aforementioned properties then it is said to be a nonparaxial pulse (beam). One example for the onset of such nonparaxiality is the ultra-narrow beam propagation in nonlinear waveguides \cite{lax}. In this situation, one has to include the spatial dispersion  which has been ignored under the SVEA and such a nonparaxiality becomes dominant in the case of miniaturized photonic devices \cite{chamorro}.

The idea of nonparaxiality is burgeoning by the early seminal work \cite{lax} of Lax \emph{et al.}, where the NLH equation has been derived by expanding field components as a power series in terms of a ratio of the beam diameter to the diffraction length ($\equiv \lambda/w_{0}$). Along similar lines, the pioneering works of Posada \textit{et al}. have dealt with the study of propagation of nonparaxial solitons in miniaturized nonlinear optical devices on the nanoscale \cite{chamorro} and the interaction dynamics of spatial Kerr solitons in the arbitrary angles of reference coordinates \cite{chamorro1}. The exact bright and dark solitons of the NLH system have been obtained for the anomalous and normal dispersion regimes in the Kerr nonlinear media along with a stability analysis \cite{christianjm}. Split-step Crank-Nicolson method has also been developed in \cite{malakuti} to perform the direct numerical simulation of the NLH system and to investigate the evolution of numerical soliton-like solutions. Recently, Lie point symmetries of the NLH system are theoretically constructed and the symmetry reductions of the NLH system are shown to be integrable, where special group of invariant solutions are also reported \cite{sakkara}. Further, exact fundamental solitons have been studied for relativistic and pseudo-relativistic formulation of the scalar nonlinear Helmholtz equation with different types of nonlinear media including cubic nonlinearity \cite{relativistic}, quintic nonlinearity \cite{Kotsmapa} and saturable nonlinearity \cite{Kotsmapa1}. Subsequently, bright and dark solitons have been studied for the Helmholtz-Manakov system, a two- component nonlinear system, for both focusing and defocusing type nonlinearities \cite{christ3}. In Ref. \cite{blair}, scalar and vector higher-order CQNLH systems  have been considered and some special soliton solutions have also been obtained. Recently, our study \cite{tamil} on the periodic waves and their limiting forms (hyperbolic solutions) of the coupled nonlinear Helmholtz (CNLH) system has explored the effect of nonparaxiality on speed, pulse-width and amplitude of the nonlinear waves in detail. A very recent work \cite{nithyandan} deals with a higher dimensional Helmholtz-Manakov system, where bright and dark solitary waves are reported. Noticing the lack of study on the CQNLH and inspired by the above interesting findings, in this paper we consider the following dimensionless cubic-quintic nonlinear Helmholtz (CQNLH) equation
\bea\label{CQNLH}
iq_{z}+\kappa~q_{zz}+\frac{s}{2}~q_{tt}+(\alpha |q|^2 +\beta |q|^{4})q=0,\,
\eea
where the evolution variables $z$ and $t$ are dimensionless longitudinal and transverse coordinates, respectively and the term $s=\pm 1$, indicates group velocity dispersion (GVD) for anomalous and normal dispersions, accordingly. In Eq. \eqref{CQNLH}, the nonparaxial parameter $\kappa=\left(1/k^{2}w_{0}\right)>0$ (the propagation constant and width of the pulse are indicated as $k$ and $\omega_{0}$), and its value ranges from $10^{-2}$ to $10^{-4}$~~~\cite{chamorro,chamorro1,christianjm,relativistic,Kotsmapa,Kotsmapa1} which is sufficient to describe the nonparaxial beam propagation. Here we keep '$\kappa$' explicitly in Eq. (1) following the standard convention and to compare our analysis on the CQNLH system with that of CQNLS system. $\alpha$ corresponds to self-phase modulation (SPM) that can be either positive or negative, which will be chosen as $\pm 1$ for convenience. The quintic nonlinearity $\beta =-E_{0}^{2}n_{4}/n_{2}$ (where $E_{0}$ is in the units of electric field, $n_{2}$ and $n_{4}$ are nonlinear refractive indices correspond to cubic and quintic nonlinearities, respectively) parameterizes the ratio of the quintic to the cubic nonlinear phases for the input pulse, which can also be either positive or negative. Thus for the system (\ref{CQNLH}), there are two possible choices for the competing nonlinearities, namely ($\alpha>0$ and $\beta<0$) and ($\alpha<0$ and $\beta>0$)  \cite{christ,christ1}.

It should be noted that in nonlinear optics the study of chirped solitons, where the frequency varies across the pulse for an instantaneous time response, has been receiving tremendous attentions from the seminal work of Hmurcik \textit{et al.},  \cite{Hmurcik}.  Later, the chirped soliton has been studied for the fundamental NLS equation and it has been shown that strength of the chirp leads to splitting of solitons \cite{desaix}. The important qualitative feature of the chirping is that compressing and amplifying solitary pulses in optical fiber and it can have ramifications in nonlinear optical fiber amplifiers, optical fiber compressors, and long-haul links with distributed dispersion loss-managed chirped solitons \cite{kruglov}. Such chirped solitons have also been studied in higher order NLS equation describing the propagation of femto second pulse in optical fibers \cite{neuer}. These studies lead to a flurry of research activities on the chirped solitons for different higher order NLS family of equations in order to explore the propagation characteristics of ultra short pulses \cite{Triki}. Apart from these analyses, the concept of frequency chirping property has also extended to different types of solitons such as chirped dissipative soliton \cite{kalashniov} and chirped Peregrine soliton \cite{chen}. However, all these studies emphasizing chirped solitons have been restricted only to the paraxial regime and the study of such chirped solitons still remains unexplored in the nonparaxial regime. To the best of our knowledge, we, for the first time, employ the idea of chirped solitons to the nonparaxial regime by considering the dimensionless CQNLH.

The paper is structured as follows. We provide MI analysis of the CQNLH system in Sec. \ref{sec2}, where we emphasize the influence of nonparaxiality on the instability growth rate. The MI dynamics of the CQNLH system is then numerically investigated by employing modified split-step Fourier (m-SSF) method, which is presented in Sec. \ref{sec3}.  In Sec. \ref{sec4}, by developing the direct integration method, we obtain exact elliptic waves, dark/gray and anti-dark solitary waves as their limiting cases for different types of nonlinearities. Also, the soliton properties and chirping are discussed in detail along with the numerical corroborations. Finally, we summarize our results in Sec. \ref{sec5}.
\vspace{-0.1cm}
\section{Modulational instability of the CQNLH system}
\label{sec2}
The MI of a plane wave in the CQNLS system has previously been studied by using both analytical and numerical methods \cite{kofane} and also by including the higher order dispersion\cite{hong}. In all these works, the dynamics of MI has been explored in the paraxial regime alone. However, the fundamental process of MI remains still unexplored in the nonparaxial regime except for very few recent studies \cite{christianjm,Kotsmapa} that just provide a glimpse on MI in the nonparaxial regime. Hence our aim is to present a complete and rigorous investigation of MI through a systematic linear stability analysis and thereby addressing the role of nonparaxial parameter on the stability of continuous wave (CW) background for the CQNLH system. We begin our analysis by assuming the following steady-state anstatz
 \bea\label{CWsol}
q(z,0)=\sqrt{P} e^{i\hat{K} z},
\eea
where $P$ is the input power and the propagation constant is read as, $\hat{K}=(-1\pm\sqrt{1+4\kappa(\alpha P+\beta P^{2})})/2\kappa$. It is quite clear that phase of the steady-state solution depends upon the intensity ($I=P$). Next, we perturb the amplitude of the steady-state solution (\ref{CWsol}) by introducing an infinitesimal  perturbation $a(z,t)$ as\bea\label{CWsol1}
q(z,t)=(\sqrt{P}+a(z,t))\exp(i\hat{K}z).
\eea
where the small perturbation $a(z,t)$ satisfies the condition $|a(z,t)|^{2}\ll P$. By substituting Eq. (\ref{CWsol1}) into Eq. (\ref{CQNLH}) and upon linearization, we get the following evolution equation for the perturbation
\begin{widetext}\bea\label{PertEq}
\pm i \frac{\partial a}{\partial z} \sqrt{1+4\kappa(\alpha P+\beta P^{2})} +\kappa \frac{\partial^{2}a}{\partial z^{2}}+\frac{s}{2}\frac{\partial^{2}a}{\partial t^{2}}+ (a+a^{*}) (P \alpha+ 2 P^{2} \beta)=0.
\eea
\end{widetext}
Let us assume the following plane wave ansatz constituting two side band components for $a(z,t)$
\bea\label{PertEq1}
a(z,t)=a_{1} \exp(i(K z- \Omega t))+a_{2} \exp(-i(K z- \Omega t)),\,\,\,\,\,\,
\eea
where $a_{1,2}$ are arbitrary real constants. Further, $K$ and $\Omega$ are complex wave number and frequency of the perturbation, respectively. A straightforward substitution of Eq. \eqref{PertEq1} into Eq. \eqref{PertEq} results in the following dispersion relation:
\bea\label{wavenumber}
K^{4}+\frac{1}{\kappa^{2}} \tilde{a}~K^2+\frac{s~\Omega^{2}}{4~\kappa^{2}}\tilde{b}=0,
\eea
where, $\tilde{a}=\left(-1- 6~\alpha~\kappa~P-8~\beta~\kappa~P^{2}+s~\kappa~\Omega^{2}\right)$ and $\tilde{b}=\left(-4~\alpha~P-8~\beta~P^{2}+s~\Omega^{2}\right)$. Here our main aim is to investigate the effect of nonparaxiality on the MI and so we consider the above general dispersion relation in the following analysis. Note that the quartic equation (\ref{wavenumber}) can admit four roots, which in general include two real roots and a pair of complex conjugate roots. It is quite apparent that the two real roots do not lead to any modulational growth while the remaining two complex conjugate pair roots can result in the MI since the largest part of  imaginary roots can result in an exponential growth for the applied perturbation. The four roots of  Eq. (\ref{wavenumber}) are then found to be
\bes\label{dispersion}\bea
K_{1-4}=\pm_{\varepsilon}\left(\frac{sgn(p_{1})|p_{1}|\pm_{\eta}\sqrt{sgn(p_{2})|p_{2}|}}{2\kappa^{2}}\right)^{\frac{1}{2}},
\eea
where the four roots $K_{1}, K_{2}, K_{3}$ and $K_{4}$ are noted through different combinations of $\varepsilon$ and $\eta$ as $(+_{\varepsilon},+_{\eta})$, $(+_{\varepsilon},-_{\eta})$, $(-_{\varepsilon},+_{\eta})$, $(-_{\varepsilon},-_{\eta})$, respectively. The expressions of $p_{1}$ and $p_{2}$ are given by
\bea
p_{1}&=&-\tilde{a},\\
p_{2}&=&-s~\kappa^{2}~\Omega^{2}\tilde{b}+\tilde{a}^{2}.
\eea
A careful analysis of relation \eqref{dispersion} shows that the roots become imaginary for the following choices, (i) $sgn(p_{1})>0$, $sgn(p_{2})>0$ with $\sqrt{|p_{2}|}>|p_{1}|$, and (ii) $sgn(p_{1})<0$, $sgn(p_{2})>0$, which make $K_{2}$ and $K_{4}$ to be complex roots. Note that we consider only the first choice with the anomalous dispersion regime for which the gain spectrum of MI is obtained as
\bea\label{gain}
G(\Omega)=2|Im(K)|=\frac{\sqrt{2}}{\kappa} \left|\sqrt{|\left(\sqrt{|p_{2}|}-|p_{1}|\right)|}\right|.
\eea\ees
\vspace{-0.035cm}
The condition $\sqrt{|p_{2}|}>|p_{1}|$ in choice (i) also requires $4 \alpha P+ 8\beta P^{2}-s\Omega^{2}>0$, which suggests the possibility of MI in the anomalous dispersion $(s=1)$ regime for $\alpha>0$ and $\beta>0$. For competing nonlinearities, $\alpha>0$ and $\beta<0$ ($\alpha<0$ and $\beta>0$) the first choice will lead to MI, if the power is assigned as, $P<\alpha/2|\beta|\left(>|\alpha|/2\beta\right)$. Otherwise, the system remains modulationally stable including the choice of defocusing cubic and quintic nonlinearities ($\alpha<0$ and $\beta <0$) in the anomalous dispersion regime ($s=1$).  An important point follows from Eq. \eqref{gain} is that the gain is inversely proportional to the nonparaxial parameter. This clearly indicates that the MI growth can be reduced by increasing the nonparaxial parameter ($\kappa$). Here to provide further insight, we separately address the influence of power, cubic as well as quintic nonlinearities and the nonparaxial parameter (spatial dispersion) on the MI dynamics.

\subsection{MI gain spectra for distinct power values}
In this subsection, we discuss the behaviour of MI in the CQNLH system (\ref{CQNLH}) by varying the power $(P)$ as well as frequency $\Omega$ in the anomalous dispersion regime with positive cubic and quintic nonlinearities  ($\alpha$, $\beta>0$). To facilitate the understanding, we start our discussion with the conventional CQNLS (i.e., CQNLH system \eqref{CQNLH} in the absence of nonparaxial parameter ($\kappa=0$)). In this case, one can observe that the CW  experiences an instability along with the usual dynamics where one notices that an increase in the power not only increases gain spectra but also extends the range of bandwidth as shown in Fig. 1(a).

Next, to have a comprehensive picture of nonparaxiality on the gain spectra, we retain the same parameters as used in Fig. 1(a)  with $\kappa=0.05$. As can be seen in Fig. ~1(b), the system admits an interesting MI behaviour wherein the usual symmetric MI gain spectrum appearing on either sides of the central frequency ($\Omega=0$) is diminished followed by a linearly increasing growth rate of the gain spectrum found at higher perturbation frequency. Note that such a spectrum features an infinite gain if one increases the perturbation frequency which resembles an infinite blow-up growth rate and we term this band as a nonparaxial spectrum. A further comparison with Fig. 1(a) indicates that the inclusion of nonparaxial parameter results in decrease in growth rate of the central MI gain spectrum. Thus the nonparaxial parameter acts as a perturbation  whose effect is to suppress the central gain and leads to a nontrivial monotonic increase in the gain spectrum near the tails of the MI spectra.

\begin{figure}[t]
\includegraphics[width=0.7\linewidth]{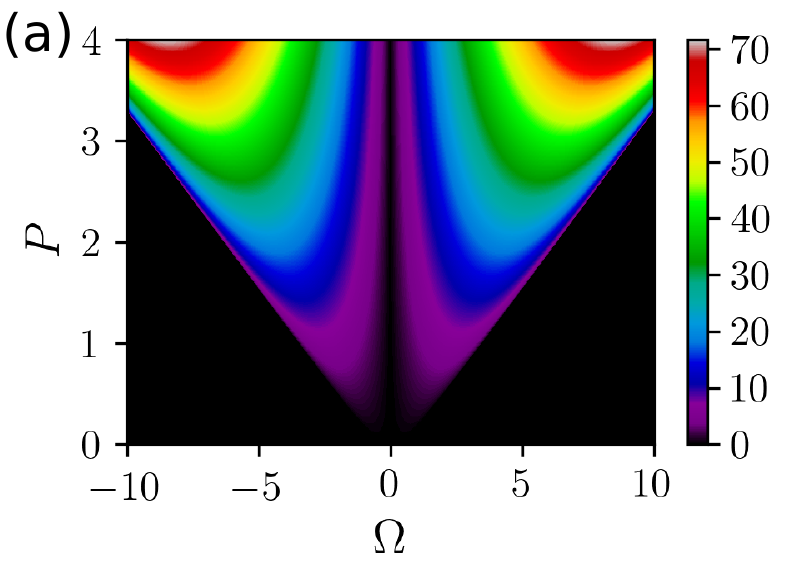}
\includegraphics[width=0.7\linewidth]{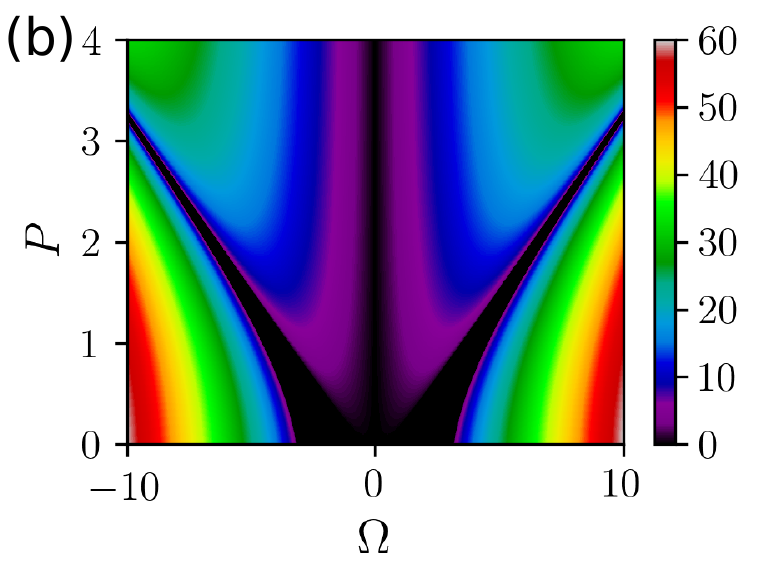}
\caption{MI gain spectra as a function of power and frequency for the CQNLH system \eqref{CQNLH} for (a) $\kappa=0$ and (b) $\kappa=0.05$, respectively. In both figures $s=1$, and $\alpha=\beta=1$.}
\end{figure}

\subsection{Role of nonlinear parameters on MI gain spectra}
 We now analyze the impacts of different types of nonlinearities in the proposed system. The plot shown in Fig. 2(a) illustrates an exclusive role of the cubic nonlinearity ($\alpha>0$) on MI for a given frequency ($\Omega$) range. We note that the gain spectrum as well as the bandwidth are increased as the cubic nonlinearity is increased. Similarly, one can notice that the quintic nonlinearity $\beta$ (see Fig. 2(b)) also exhibits the same ramifications as observed in the Fig. 2(a), where the gain and bandwidth are increased as the quintic nonlinearity is hiked up.
\begin{figure}[t]
\includegraphics[width=0.49\linewidth]{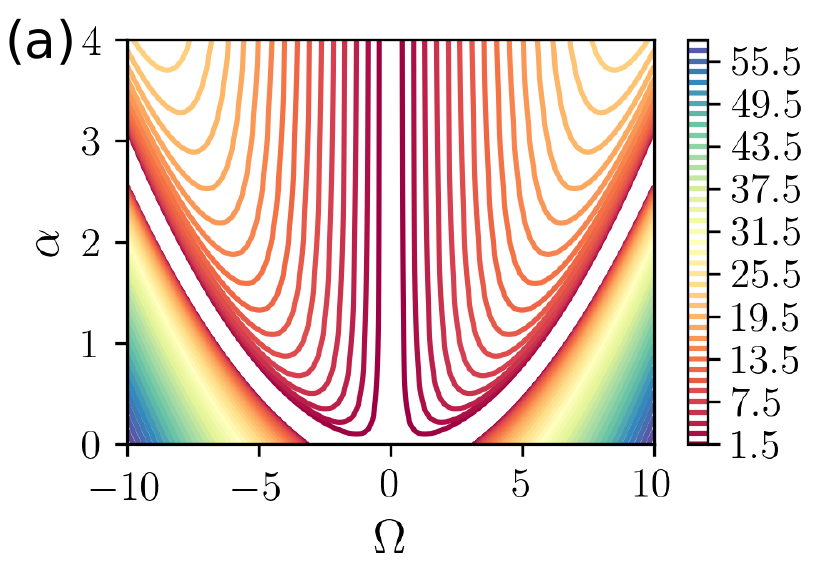}
\includegraphics[width=0.49\linewidth]{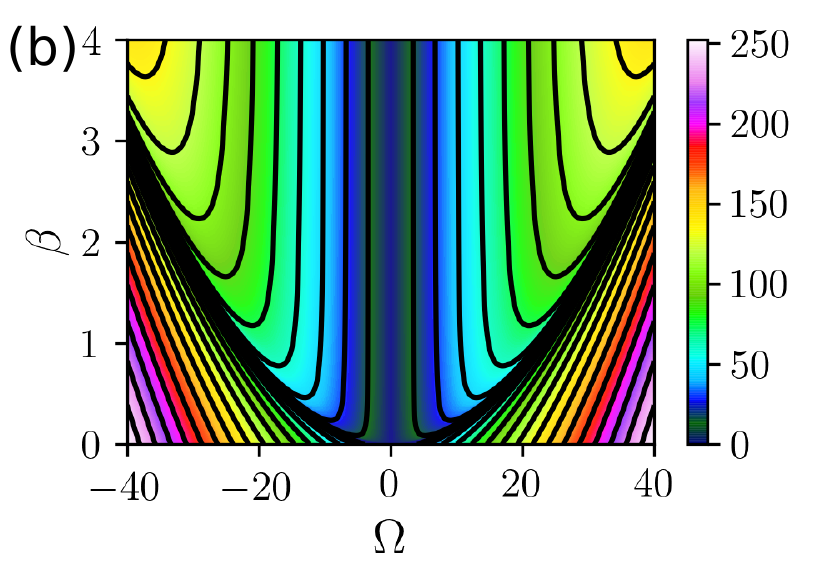}
\caption{Plots showing  MI gain spectra  (a) as a function of $\alpha$ and $\Omega$ for $\beta=0$ and (b) as a function of $\beta$ and $\Omega$ for $\alpha=0$. In both figures $s=1$, $P=8$, and $\kappa=0.05$.}
\end{figure}

\subsection{MI gain spectra for variation of nonparaxial parameter}
Here, we discuss the impact of nonparaxial parameter ($\kappa$), which accounts for the spatial dispersion, on MI gain spectrum of the system (\ref{CQNLH}). In this regard,  we fix the power, cubic $(\alpha)$ and quintic $(\beta)$ nonlinear co-efficients. Figures~3(a) and 3(b) illustrate the MI behaviour in the anomalous dispersion regime. The MI spectrum behaves as that of the conventional CQNLS system when $\kappa=0$. However, as $\kappa$ increases, there is a nontrivial monotonic increase in the gain spectrum near the tails of the conventional MI band. At this particular point, the perturbation of the CW is influenced significantly by the spatial dispersion and the gain increases monotonically. Also, one can note that the separation distance between the conventional sideband and the nonparaxial spectra  decreases as the nonparaxial parameter $\kappa$ increases, which further means that the tail distortion occurs quickly as the nonparaxiality is increased.

\begin{figure}[t]
	\includegraphics[width=0.49\linewidth]{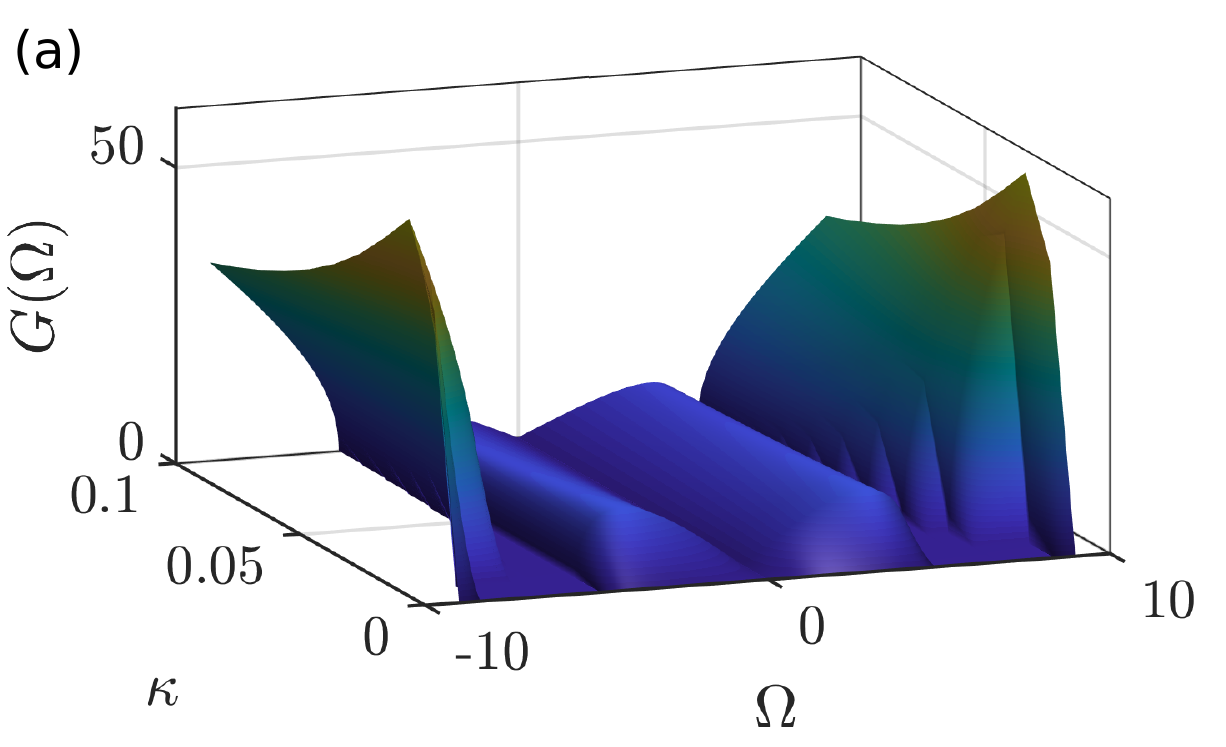}
	\includegraphics[width=0.49\linewidth]{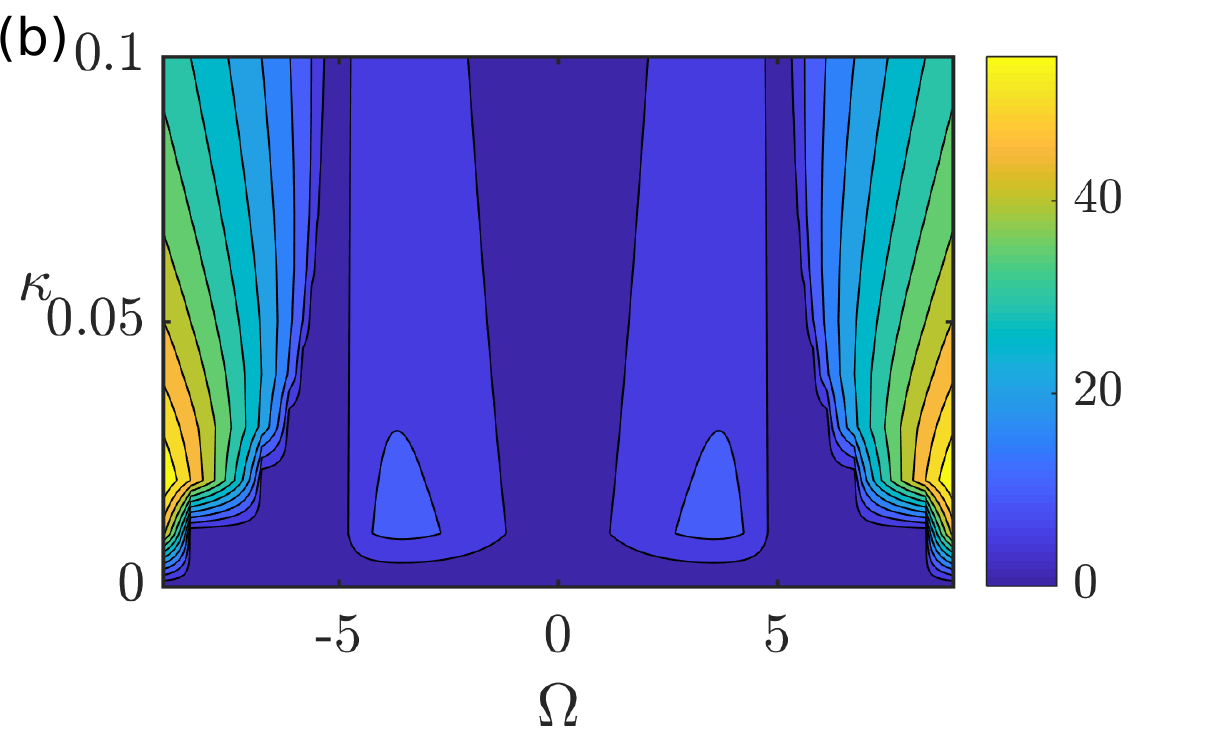}\\
	\includegraphics[width=0.49\linewidth]{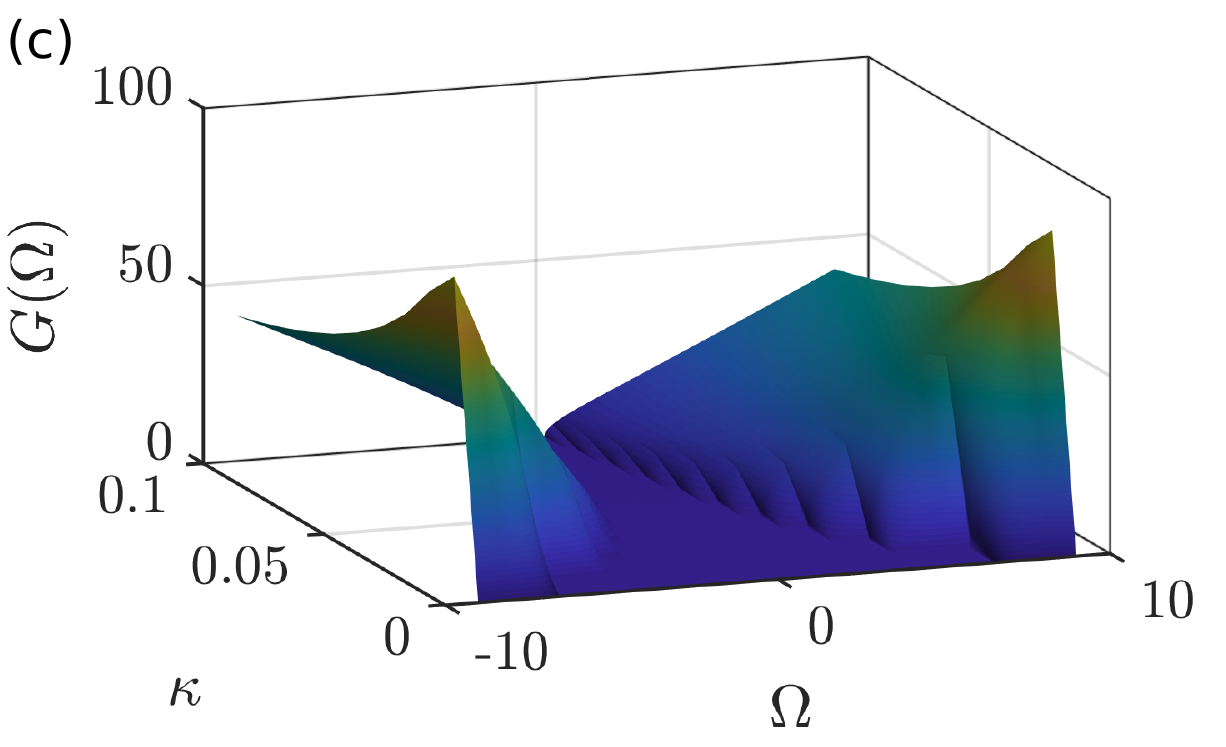}
	\includegraphics[width=0.49\linewidth]{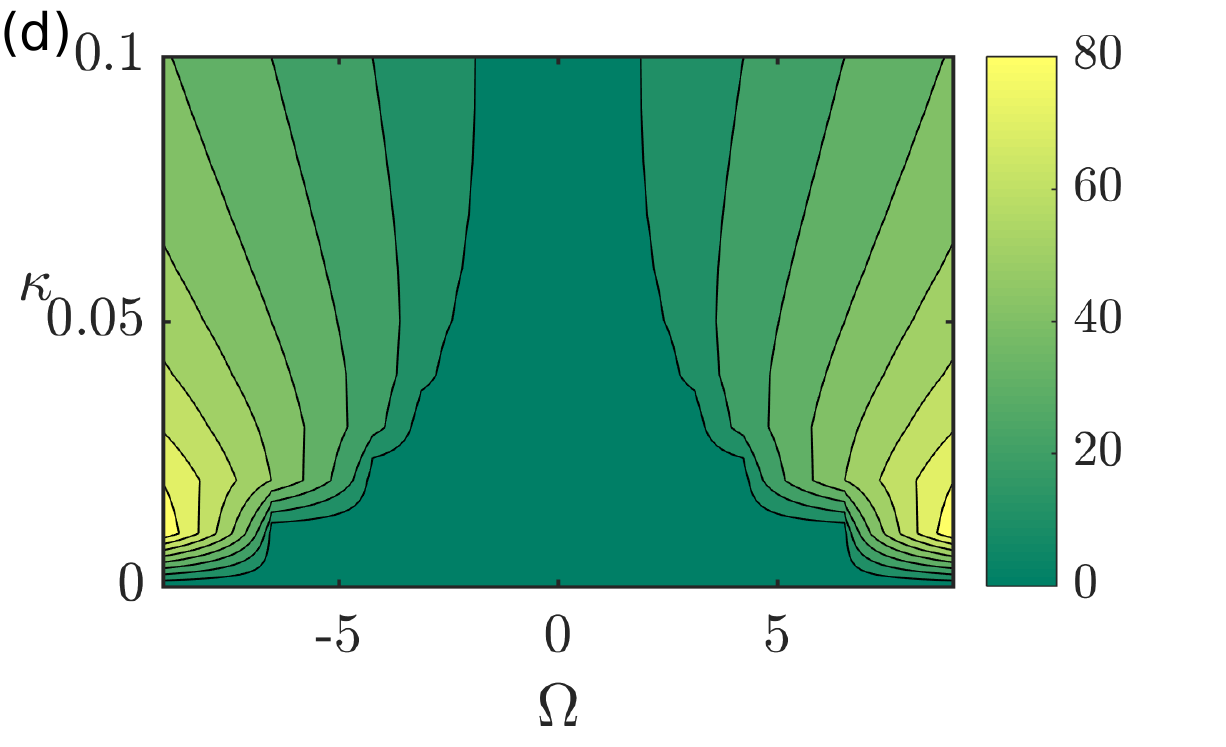}\\
	\includegraphics[width=0.46\linewidth]{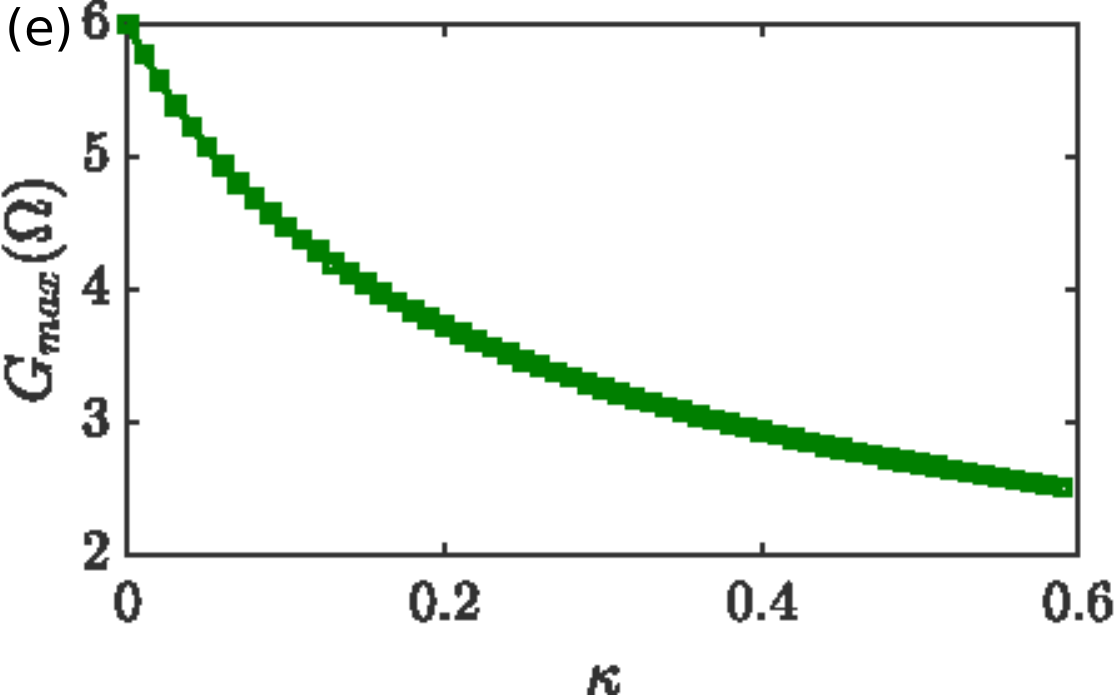}
	\includegraphics[width=0.46\linewidth]{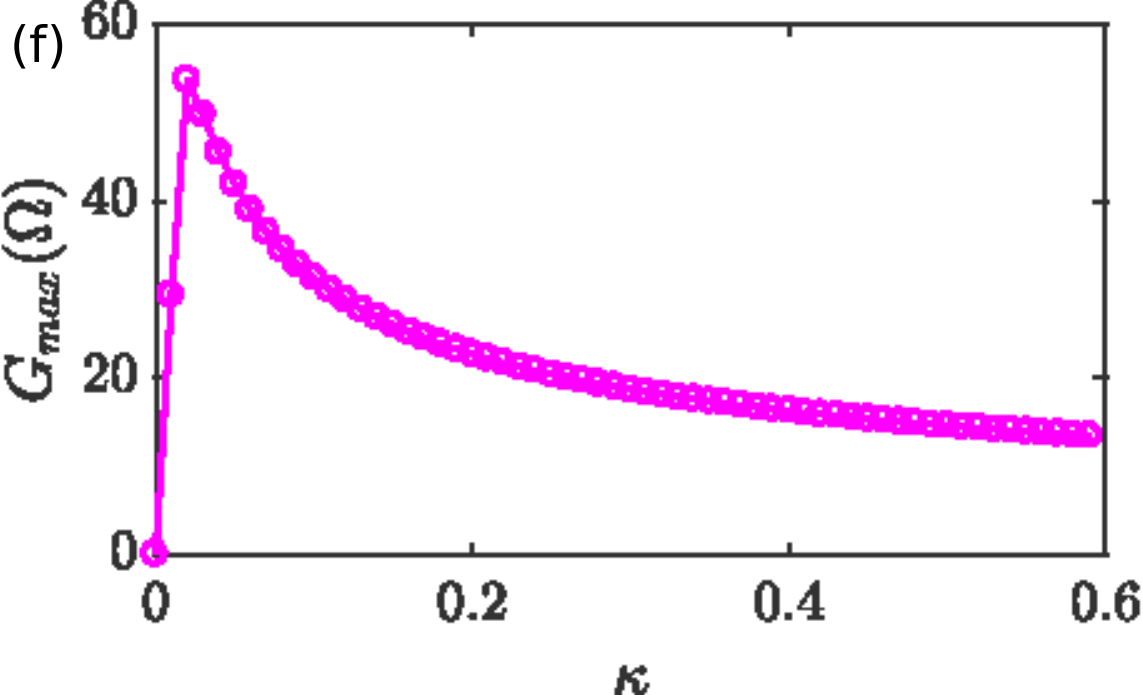}\\
\caption{MI gain spectra for the variation of  frequency and nonparaxial parameter in the CQNLH system \eqref{CQNLH}. In top panels we fix $\beta=1$ whereas the quintic nonlinearity is fixed to be $\beta=-1$ in the middle panels. The growth rate of conventional MI band and that of monotonically growing spectrum (nonparaxial spectra) plotted against the nonparaxial parameter $(\kappa)$ for the choice ${\bf\alpha=\bf\beta=1}$ and $P=1.5$ are shown in the bottom panels (e) and (f), respectively. The parameters are $s=\alpha=1$ and $P=1.5$.}
\end{figure}
\begin{figure}[t]
	\includegraphics[width=0.49\linewidth]{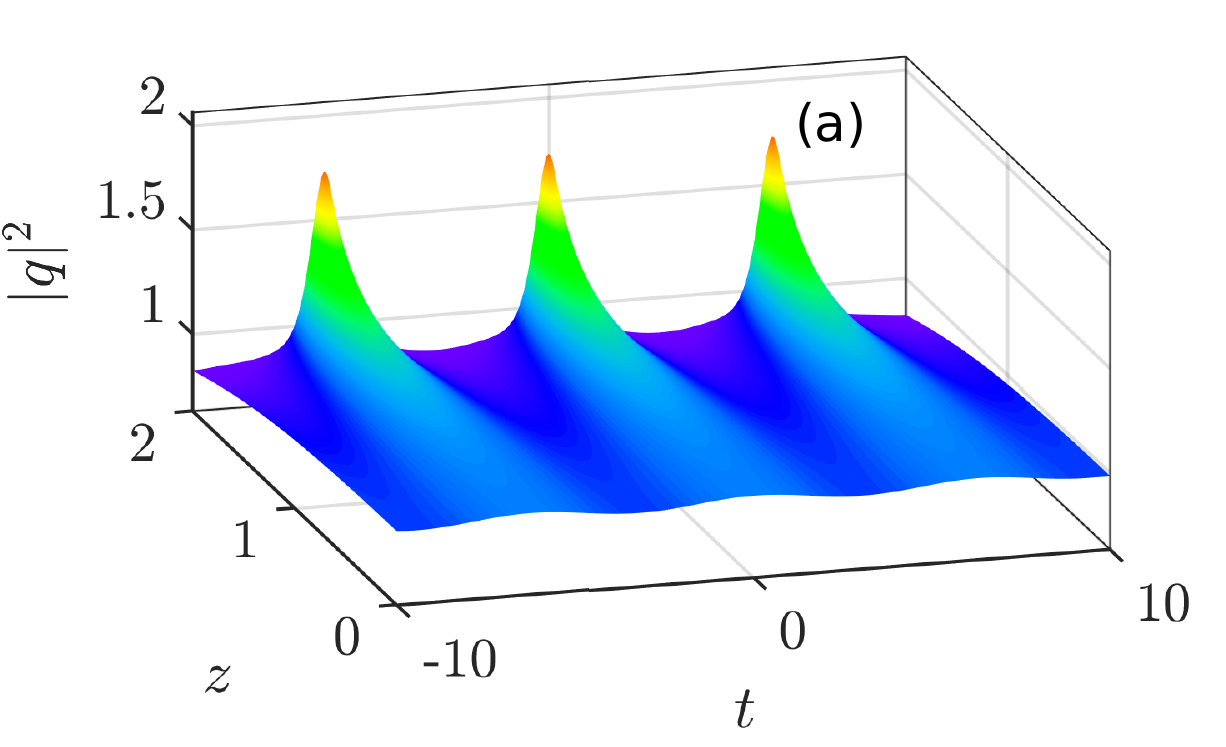}
	\includegraphics[width=0.49\linewidth]{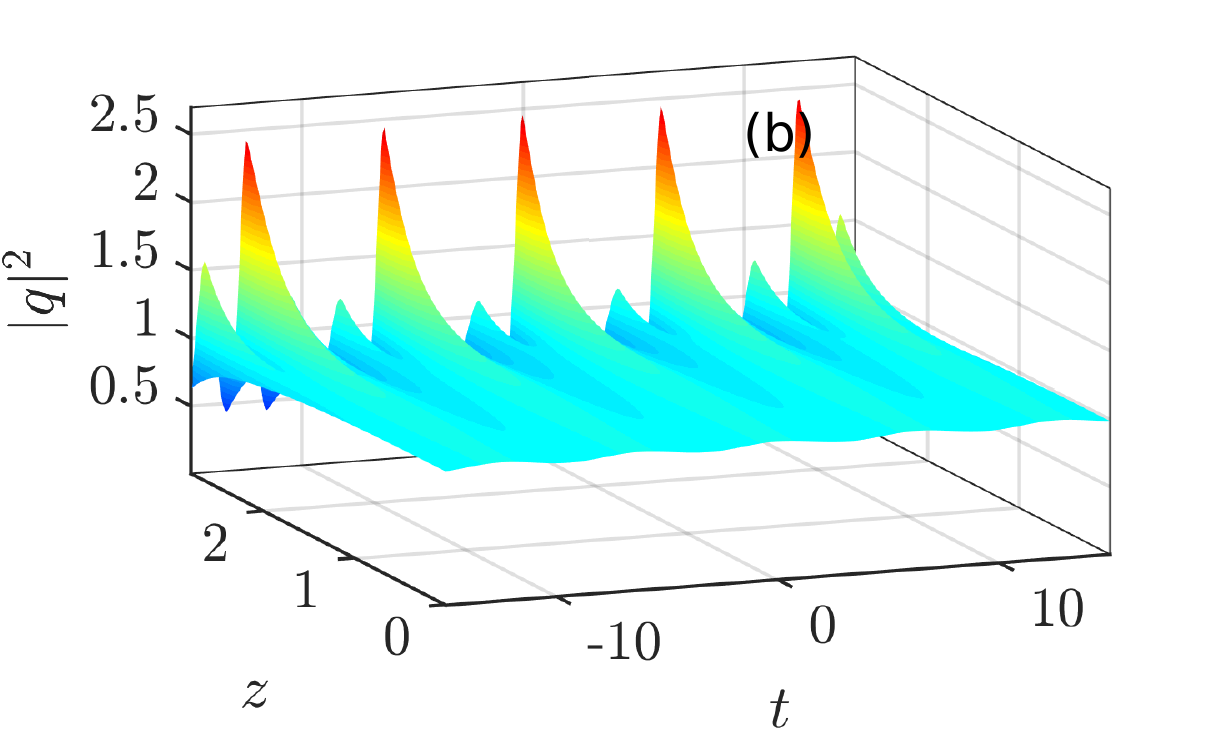}\\
	\includegraphics[width=0.49\linewidth]{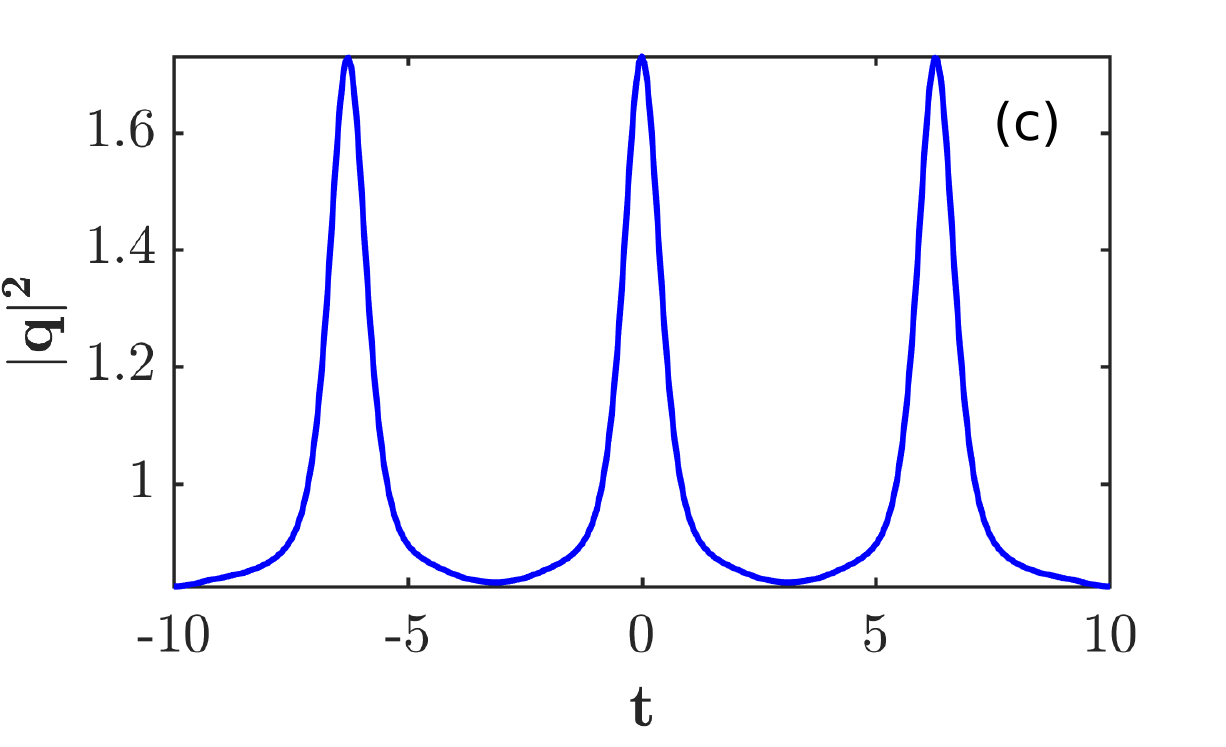}
	\includegraphics[width=0.49\linewidth]{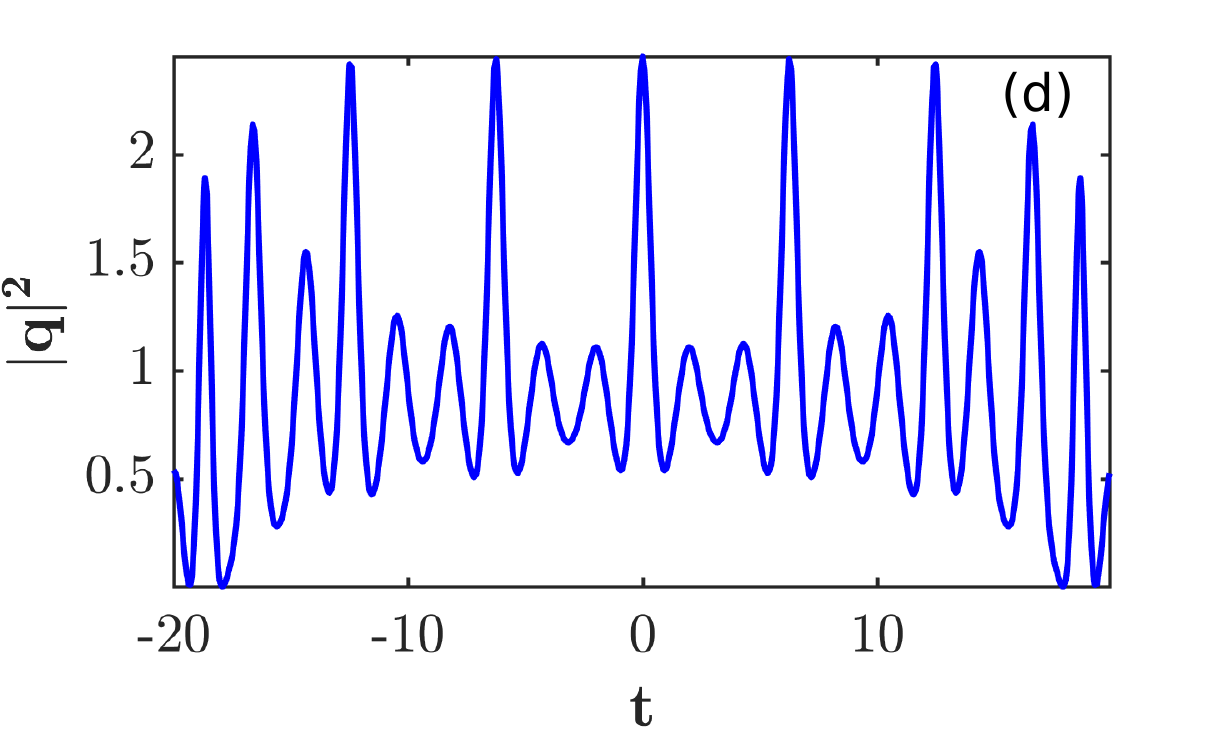}\\
	\includegraphics[width=0.49\linewidth]{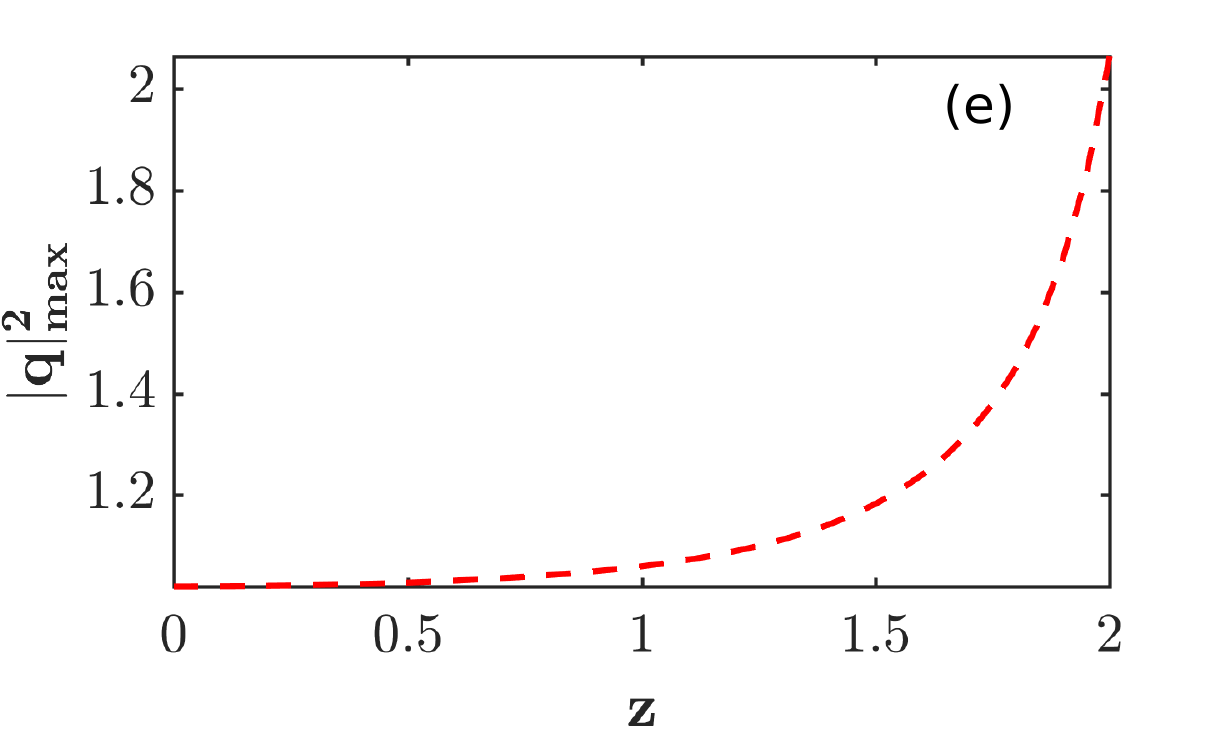}
	\includegraphics[width=0.49\linewidth]{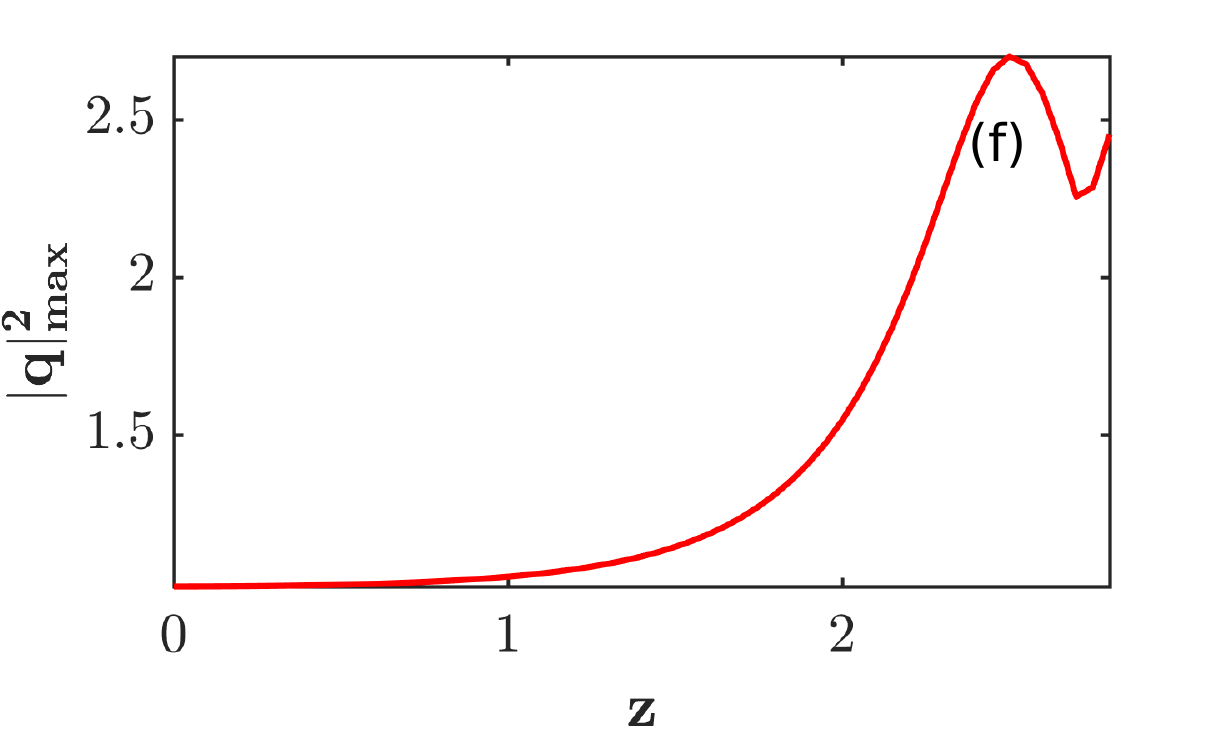}\\
	\caption{(a, b) Evolution of perturbed CW and (e, f) the growth rate for CQNLH system \eqref{CQNLH}  when $\kappa=0$ (left panels) and $\kappa=0.05$ (right panels). The parameters are assigned such that, $s=\alpha=\beta=1$. Note that middle panels show the train of ultra-short pulses at $z=2$.}
\end{figure}
We would like to remark here that the role of defocusing nonlinearity on MI spectra has already been studied in many CQNLS systems and it is found that the former parameter results in complete stabilization and thereby not allowing any MI spectral bands. However, this scenario is slightly different in the nonparaxial system and it is depicted in Figs. 3(c) and 3(d).  To elucidate this phenomenon,  we plot the MI spectrum for the defocusing qunitic nonlinearity ($\beta<0$) with fixing all other parameters same as in Figs. 3(a) and 3(b). Here, one observes that although the MI is absent in and nearby regions of central perturbation frequency, there is a monotonic increase in the gain spectra at the higher detuning frequency which hints a trademark signature of nonparaxial parameter even in such competing CQNLH systems. To identify the impact of nonparaxial parameter in detail, we show the behaviour of maximum MI growth rate of conventional MI symmetric bands and the monotonically growing spectrum found at the higher detuning frequency as $\kappa$ varies, separately in Figs. 3(e) and 3(f) respectively. As seen in Fig. 3(e), it is quite clear that the nonparaxial parameter results in decrease of MI spectrum while increasing the former value, which further attests our analytical predictions of Eq. 7(d). For the monotonically increasing spectrum (see Fig. 3(f)), one can notice that the maximum MI growth rate blows-up to a certain  high value ($\approx 52$) when the nonparaxial parameter attains some critical values ($\kappa>0.05$) and beyond this value, the nonparaxial parameter leads to a suppression of the MI growth rate which gets saturated for larger values of $\kappa$. Note that in both two cases, the maximum instability growth rate is calculated for a given range of finite perturbation frequency.
\section{MI Dynamics of the CQNLH system}
\label{sec3}
To study the long time behaviour of the perturbed CW solution, we perform a numerical study of the CQNLH system \eqref{CQNLH} in the anomalous dispersion regime ($s=1$) with focusing nonlinearities ($\alpha>0$ and $\beta>0$) by employing the m-SSF method with the aid of MATLAB 2016b. For the numerical calculations, we choose $\Delta z = 0.0001$ and $\Delta t = 0.12$ with 1024 Fourier points and performed $10,000$ (sufficient) iterations. In this connection, we consider the following initial condition of CW with small periodic perturbation for the CQNLH system \eqref{CQNLH}
\bea
u(z=0,t) = A_{0} \left[1 + \epsilon \cos(\hat{\omega} t)\right].
\eea
Here, we assume the perturbation value as $\epsilon=10^{-2}$ and $\hat{\omega}$ is the perturbed wave frequency. First, we show the evolution of perturbed CW solution in the CQNLH system \eqref{CQNLH} in the absence of nonparaxial parameter ($\kappa=0$). In this case, up to distance $z=1.5$, the CW possess constant background but after that, the CW is not stable against the small perturbation and one can observe the propagation of the ultra-short pulses in periodic manner as shown in Figs. 4(a) and 4(c). This implies that the CQNLS system may have bright solitary wave solutions. It is well in agreement with the earlier analytical predictions. The growth rate of the ultra-short train pulses is  shown in Fig. 4(e), where the maximum intensity increases exponentially as the longitudinal distance is increased.

In the second part, we study the role of nonparaxial parameter on the MI. For this purpose, we perform a numerical simulation of the evolution of perturbed CW solution in the CQNLH system \eqref{CQNLH} with $\kappa\neq0$. Here, the CW background becomes unstable completely against the small perturbation, and we get the ultra short nonparaxial pulses with internal oscillations which are shown in Figs. 4(b) and 4(d). For this case the ultra-short pulses are distorted at the wings thereby increasing there width as shown in Figs. 4(b) and 4(d) which confirms the gain distortion at the wings discussed in the previous section. These are the special signatures of the nonparaxial parameter as mentioned in Ref. \cite{christ3}. However, from Fig. 4(f), we notice that the growth rate is decreased slightly after reaching the maximum.
\vspace{-0.04cm}
\section{Chirped nonlinear waves}
\label{sec4}
In this section, we pay attention to construct the chirped periodic waves and solitary waves of the system (\ref{CQNLH}). For this purpose, we consider the following chirped traveling wave solution
\vspace{-0.09cm}
\bea\label{ansatz1}
q(z,t)=\psi(\xi)~e^{i(\phi(\xi)- k z)},\quad\xi=t-u z,\quad u=\frac{1}{v}.
\eea
where the amplitude $\psi$ and the phase $\phi$ are real functions of $\xi$. The parameter $v$ describes the group velocity of the wave packet and $k$ is the wave number. The chirp is expressed as $\delta\omega=-(\partial/\partial t)(\phi(\xi)-k z)=-d \phi/d\xi$. Substituting Eq.~(\ref{ansatz1}) into Eq.~(\ref{CQNLH}) and collecting the imaginary parts, we obtain
\bea\label{im1}
(\kappa u^{2}+\frac{s}{2})\psi\frac{d^{2} \phi}{d \xi^{2}}+[2(\kappa u^{2}+\frac{s}{2})\frac{d \phi}{d \xi}-u+2 u k \kappa]\frac{d \psi}{d \xi}=0.\,\,\,\,\,\,\,\,\,\,\,
\eea
By integrating Eq. \eqref{im1} twice with respect to $\xi$, we get $\phi(\xi)=c_{2}+
\int \left(u~(1-2~\kappa~k)/(s+2~u^{2}~\kappa)+c_{1}/\psi^{2}\right)d\xi$, where $c_{1}$ and $c_{2}$ are integration constants, which result in the following chirp parameter
\bea\label{chirp}
\delta \omega=-\left(\frac{u~(1-2~\kappa~k)}{s+2~u^{2}~\kappa}+\frac{c_{1}}{\psi^{2}}\right).
\eea
According to Eq. \eqref{chirp}, the chirp parameter is characterized by velocity $u$, wave number $k$, group velocity dispersion $s$, nonparaxial parameter $\kappa$ and the intensity of the chirped pulse. The first term in Eq.~(\ref{chirp}) indicates the constant chirp is free from the intensity of the solitary waves. However, the second term shows an inverse variation of chirp with respect to the intensity of the solitary waves.  On the contrary, the chirping of higher order nonlinear system is directly proportional to the intensity of the resulting wave and it saturates at some finite value as $t\rightarrow\pm\infty$ \cite{Hmurcik,desaix}. Next, we consider the following nonlinear ordinary differential equation obtained by equating the real parts after substitution of \eqref{ansatz1} into \eqref{CQNLH}:
\begin{widetext}\bea\label{re}
\label{re1}
a \frac{d^{2} \psi}{d \xi^{2}}- \left(k(-1+k~\kappa)\psi+(-1+2~k\kappa)u\psi~\frac{d \phi}{d \xi}+\left(\kappa~u^{2}+\frac{s}{2}\right)\psi\left(\frac{d \phi}{d \xi}\right)^{2}\right)
+\alpha~\psi^{3}+\beta~\psi^{5}=0.
\eea\end{widetext}
Here, $a=\left(\kappa~u^{2}+s/2\right)$. Substitution of the chirp parameter (\ref{chirp}) into the above equation (\ref{re}) yields
\bea\label{real}
a~\frac{d^{2}\psi}{d \xi^{2}}+b~\psi-\frac{c^{2}_{1}}{\psi^{3}}+\alpha~\psi^{3}+\beta~\psi^{5}=0,
\eea
where $b=k\left(1-\kappa k\right)+u^{2}(1-2\kappa k)^{2}/(s+2u^{2}\kappa)$. After integrating Eq.~(\ref{real}) followed by a simple manipulation, we get
\bea\label{ode}
\frac{1}{4}\left(\frac{d(\psi)^{2}}{d\xi}\right)^{2}=a^{'} \psi^{2}-b^{'}~\psi^{4}-c^{'}~\psi^{6}-d^{'}\psi^{8}+a_{0}^{'},\,\,\,\,\,\,
\eea
where $a^{'}=2C$ (in which $C$ is an integration constant), $b^{'}=\left(b/a\right)$, $c^{'}=\left(\alpha/a\right)$ $d^{'}=\left(\beta/a\right)$ and $a_{0}^{'}=c_{1}^{2}$. It should be noted that in the above equation \eqref{ode} all the coefficients contain the nonparaxial parameter ($\kappa$) except the quadratic term. The presence of nonparaxial parameter makes Eq. \eqref{ode} to be unique as compared with earlier works done \cite{neuer}. The details of solving Eq. \eqref{ode} are provided in Appendix A.

In what follows, we first address the construction of chirped periodic wave for the CQNLH system with focusing cubic and defocusing quintic nonlinearities (i.e., competing nonlinearities) that features the chirped anti-dark solitary wave in the hyperbolic limit. Following that, we construct bright solitary wave for the competing nonlinearities by tuning the parameters in the anti-dark solution. The rest of the paper deals with construction of periodic wave for the CQNLH with defocusing cubic and quintic nonlinearities. Indeed, by adjusting the solution parameters of the periodic wave, one can obtain chirped gray and dark solitary waves in the hyperbolic limit. To substantiate our analytical results, we present numerical evolutions of intensity profiles corresponding to the exact analytical nonlinear waves by employing the m-SSF method as explained in the MI section.
 \begin{figure}[t]
	\includegraphics[width=0.45\linewidth]{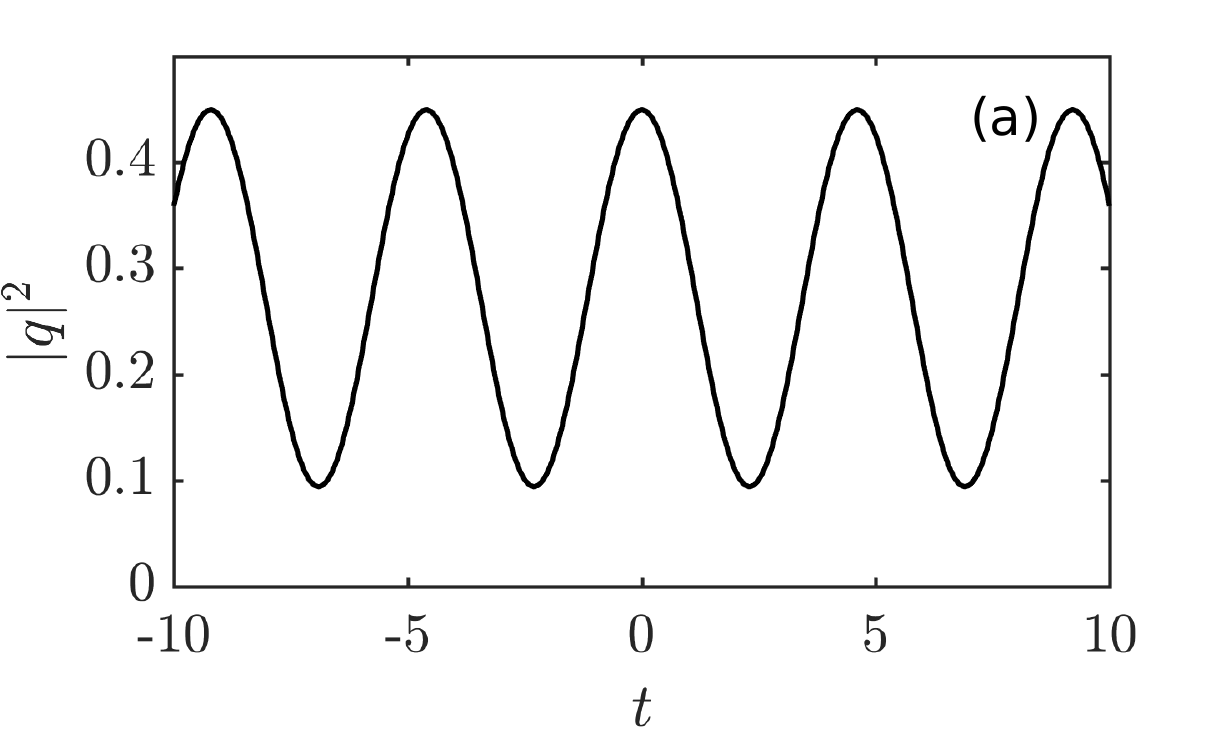}
	\includegraphics[width=0.45\linewidth]{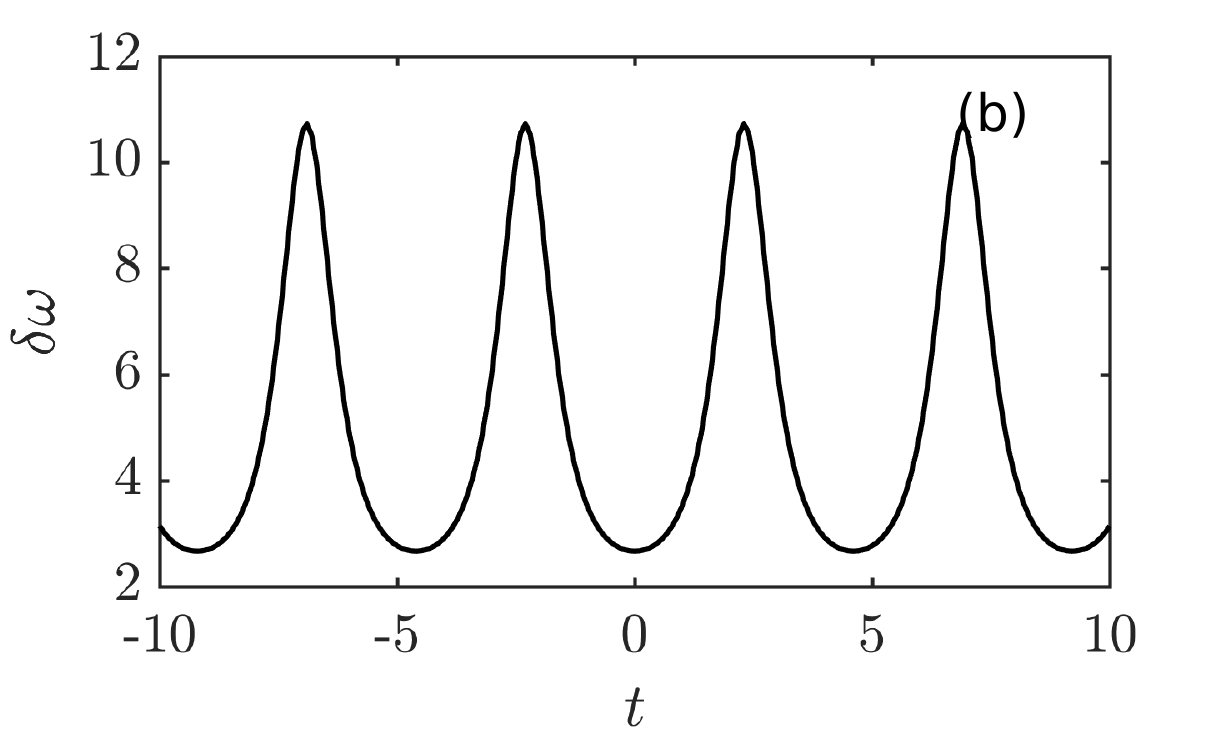}
	\includegraphics[width=0.45\linewidth]{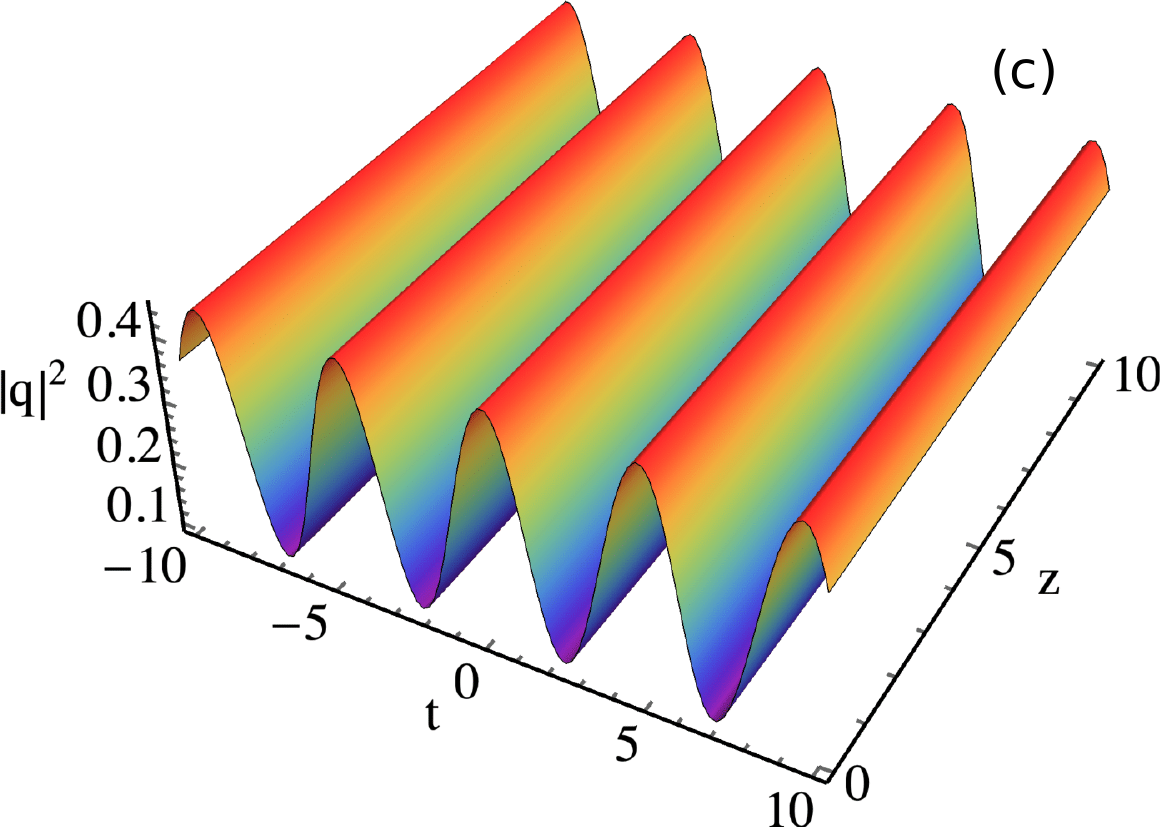}
	\includegraphics[width=0.5\linewidth]{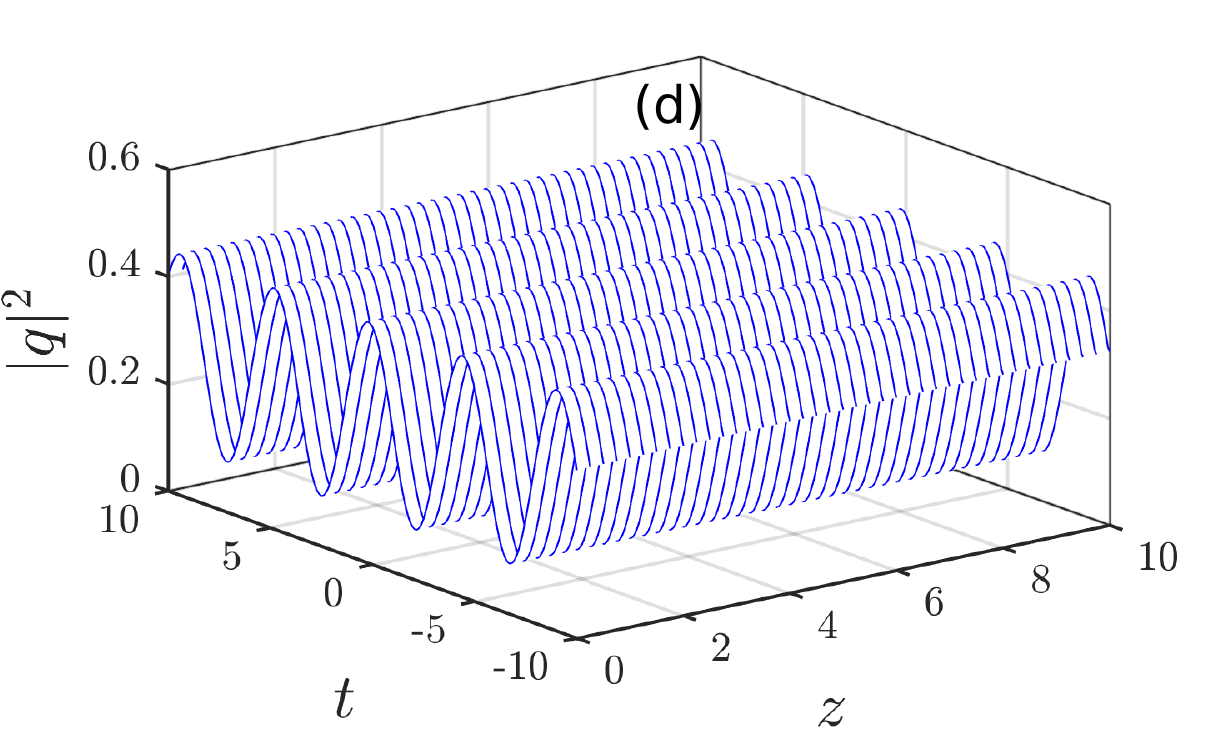}
	\caption{Here and in all forthcoming figures (except Figs. 7 and 9), top left panel refers to the two-dimensional plots of solitary/periodic waves while the top right panel indicates their corresponding two-dimensional chirping profiles.  Also, the evolution of analytically calculated profiles are shown in the bottom left panel whereas the bottom right panel shows their corresponding numerical corroborations which are in good agreement with analytical results. The parameters are assigned as, $\alpha=1$, $\beta=-1$, $\kappa=0.02$, $k=0.28$, $u=0.01$, $s=1$, $C=0.29$, $c_{1}=1$ and $\xi_{0}=0$. }
\end{figure}
\vspace{-0.2in}
\subsection{Periodic wave for competing CQNLH system with $\alpha>0$ and $\beta<0$}
One can obtain the periodic wave solution of the CQNLH system  \eqref{CQNLH} in the physical setting of the anomalous dispersion  counterbalanced by competing ($\alpha>0$ and $\beta<0$) nonlinearities by applying the direct integration method to Eq. \eqref{ode}. The periodic wave solution obtained in terms of Jacobi elliptic $\sine$ function is given below
\bea\label{solution1}
q=\left(\frac{(v_{2}-v_{1} \tilde{\alpha}^{2} \sn^{2}(\tilde{u},m))}{1-\tilde{\alpha}^{2}\sn^{2}(\tilde{u},m)}\right)^\frac{1}{2}~e^{i(\phi(\xi)- k z)}.
\eea
 Here, $\tilde{u}=\sqrt{|d^{'}| (v_{2}-v_{3}) (v_{2}-v_{4})}(\xi-\xi_{0})$ and the modulus parameter can be expressed by $m=(v_{1}-v_{4})/(v_{2}-v_{4})\tilde{\alpha}^{2}$ $\left( \text{in which}~\tilde{\alpha}^{2} =(v_{2}-v_{3})/(v_{1}-v_{3})\right)$ and $(0\leq m\leq 1)$.
One can infer from  the expressions of $m$ and $\alpha^{2}$ and from the condition on the four roots ($v_{i}, i=1,2,3,4$) that $\alpha^{2}<1$ and hence the denominator in solution (15) remains to be positive for all choice of system parameters resulting in roots fullfilling the condition $v_{1}>v_{2}>v_{3}\geq v_{4}$. The solution \eqref{solution1} is characterized by four parameters and its amplitude is determined by these four real roots ($v_{1}>v_{2}>v_{3}\geq v_{4}$). The phase is given by $\phi(\xi)- k z$, in which $\phi(\xi)$ implies the phase modulation during the propagation of periodic wave. The corresponding chirping parameter is given below

\bea\label{chirp01}
\delta\omega=-\left(\frac{u~(1-2~\kappa~k)}{s+2~u^{2}~\kappa}+\frac{c_{1}(1-\tilde{\alpha}^{2}\sn^{2}(\tilde{u},m))}{\left(v_{2}-v_{1} \tilde{\alpha}^{2} \sn^{2}(\tilde{u},m)\right)}\right).\,\,\,\,\,
\eea

For $c_1=0$, the periodic wave \eqref{solution1} can have only constant chirp, whereas the chirp is ruled out for the case of $c_{1}=u=0$. Figures 5(a) and 5(c), respectively, show the two and three dimensional plots of the periodic wave given by the exact solution \eqref{solution1} and the associated chirp  given by Eq. \eqref{chirp01} is displayed in the top right panel (Fig. 5(b)) where the chirp parameter behaves reciprocal to that of the intensity. In parallel, we demonstrate the numerical evolution of the intensity profile for the periodic wave as shown in Fig. 5(d) which well corroborates with our analytical predictions. The intensity of the periodic wave diminishes with the increase of the value of the nonparaxial parameter ($\kappa$) while the chirp of the periodic wave increases for increasing the former as shown in Fig. 6. Thus by altering chirp, the intensity can be controlled in this case.
\begin{figure}[t] 	
	\includegraphics[width=0.55\linewidth]{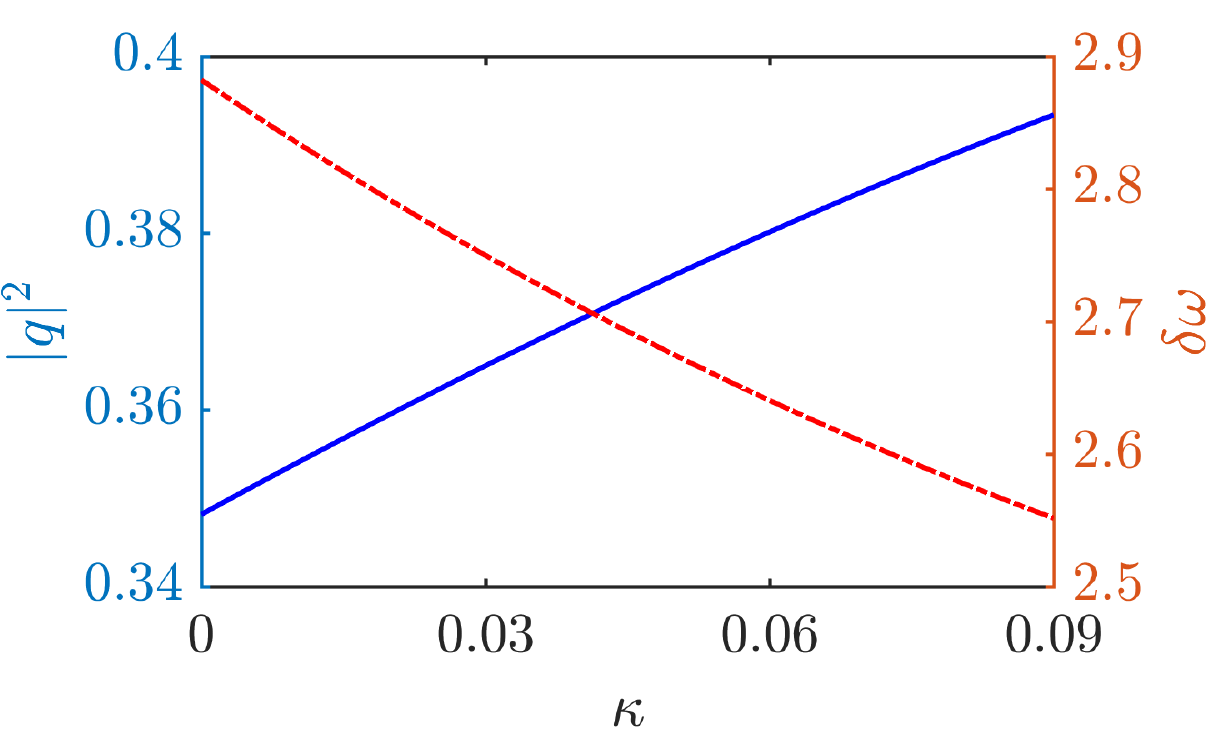}

	\caption{Variation of intensity and chirp profile as a function of nonparaxial parameter. The parameters are same as in Fig. 6}
\end{figure}
\subsection{Anti-dark and bright solitary waves ($\alpha>0$ and  $\beta<0$)}
In this sub-section, we are going to study the hyperbolic limit $(m=1)$ of the above elliptic solution which turns out to be an anti-dark wave solution of the CQNLH system \eqref{CQNLH} with focusing cubic nonlinearity $\alpha>0$ and defocusing quintic nonlinearity $\beta<0$. We arrive at the following anti-dark solitary wave which is a bright soliton on top of a non vanishing background for the condition $v_{3}=v_{4}$, $v_{1}>v_{2}$ \cite{byrd}, in \eqref{solution1}
\bea\label{focusing1}
q=\left(\frac{(v_{2}-v_{1} \tilde{\alpha}^{2} \tanh^{2}(\tilde{u}))}{1-\tilde{\alpha}^{2}\tanh^{2}(\tilde{u})}\right)^\frac{1}{2}~e^{i(\phi(\xi)- k z)}.\,\,\,\,\,
\eea
where $\tilde{u}=\sqrt{|d^{'}|} (v_{2}-v_{4})(\xi-\xi_{0})$, in which $\sqrt{|d^{'}|}(v_{2}-v_{4})$ determines the inverse width of the solitary wave, $\tilde{\alpha}^{2}=(v_{2}-v_{4})/(v_{1}-v_{4})$  and $m=1$. Here $v_{i}, i=1,2,3,4$ are dependent on system parameters $\kappa$, $s$, $\alpha$, and $\beta$. The chirping parameter associated with this anti-dark solitary wave solution reads as
\bea\label{antidarkchirp}
\delta\omega&=&-\left(\frac{u~(1-2~\kappa~k)}{s+2~u^{2}~\kappa}+\frac{c_{1}(1-\tilde{\alpha}^{2}\tanh^{2}(\tilde{u}))}{\left(v_{2}-v_{1} \tilde{\alpha}^{2} \tanh^{2}(\tilde{u})\right)}\right).\,\,\,\,\,
\eea
In the left panels, the two- and three- dimensional plots of the anti-dark solitary waves  given by \eqref{focusing1} are displayed, respectively, in Figs.~7(a) and 7(c). Their corresponding two-dimensional plot of the chirping profile is shown in Fig.~7(b), in which the chirp propagates in the negative direction. Thus, the anti-dark solitary wave does have the negative chirping as that of solution \eqref{solution1}. Figure 7(d) shows the corresponding numerical evolution of the anti-dark solitary wave. Asymptotically (i.e., $|\xi|\rightarrow \infty$) the intensity reaches a constant value given by $\left(\frac{v_{2}-v_{1} \alpha^{2}}{1-\alpha^{2}}\right)$.
\begin{figure}[ht]
\includegraphics[width=0.45\linewidth]{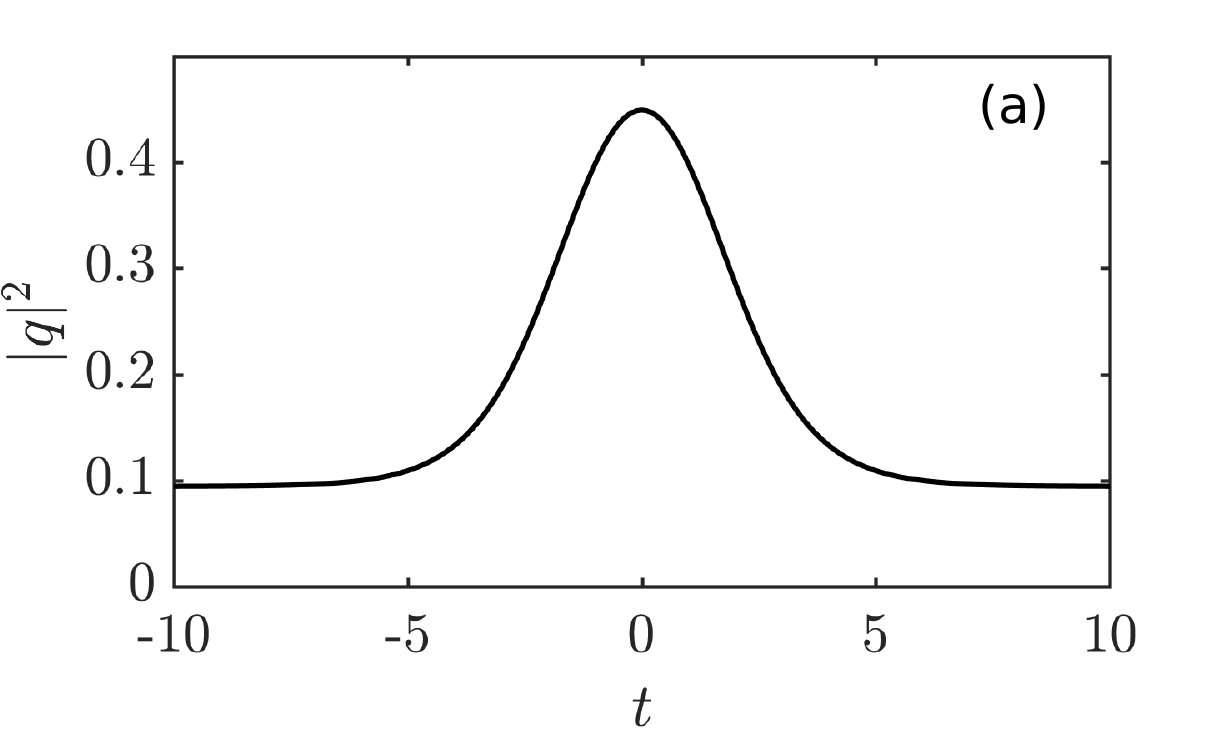}
\includegraphics[width=0.45\linewidth]{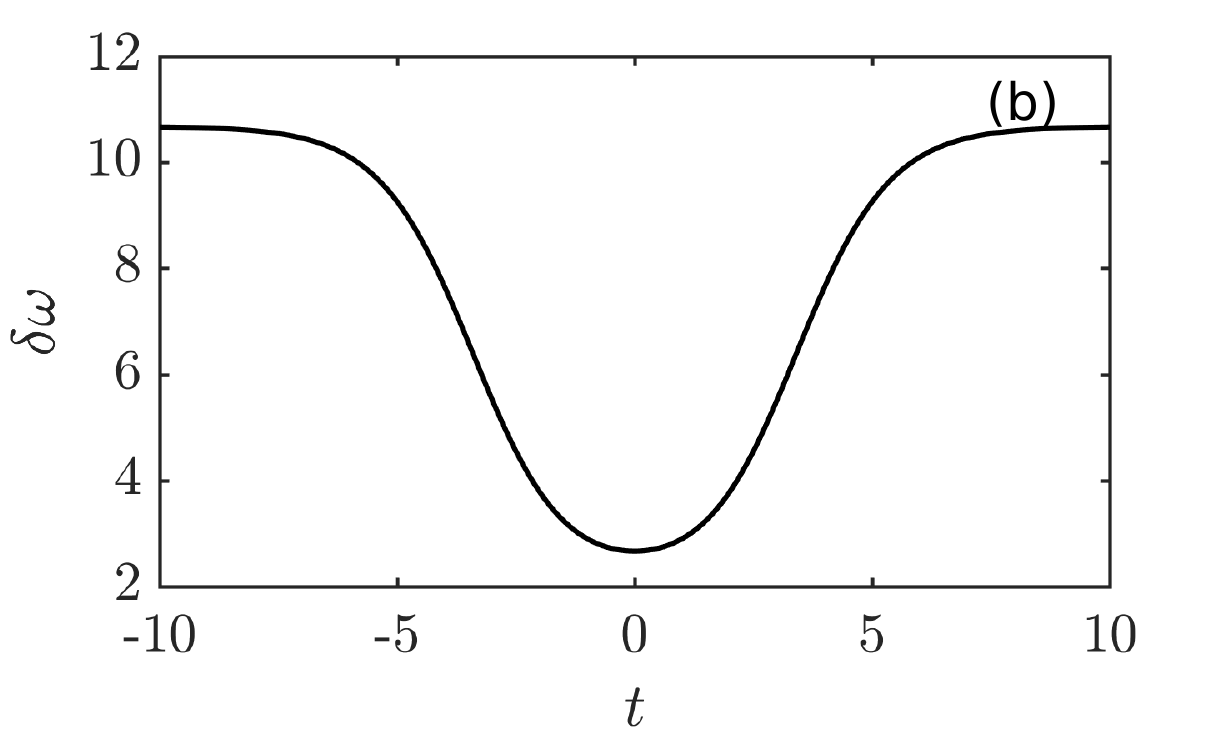}
\includegraphics[width=0.45\linewidth]{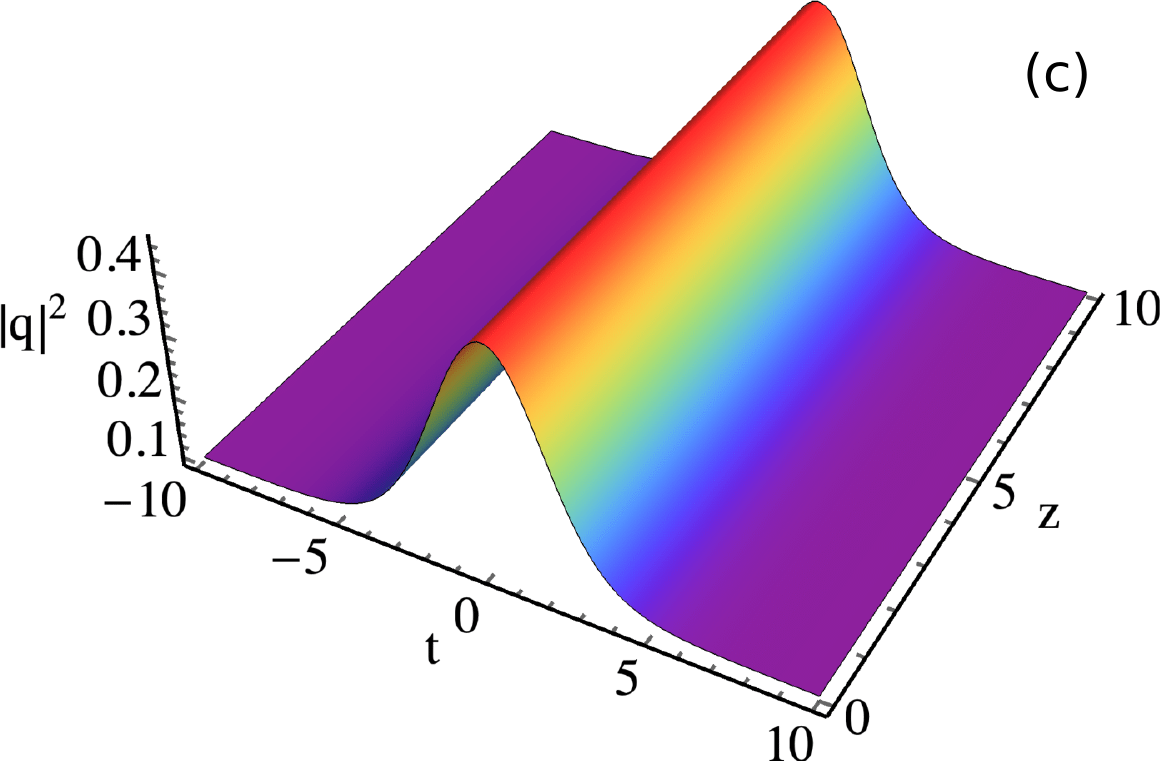}
\includegraphics[width=0.5\linewidth]{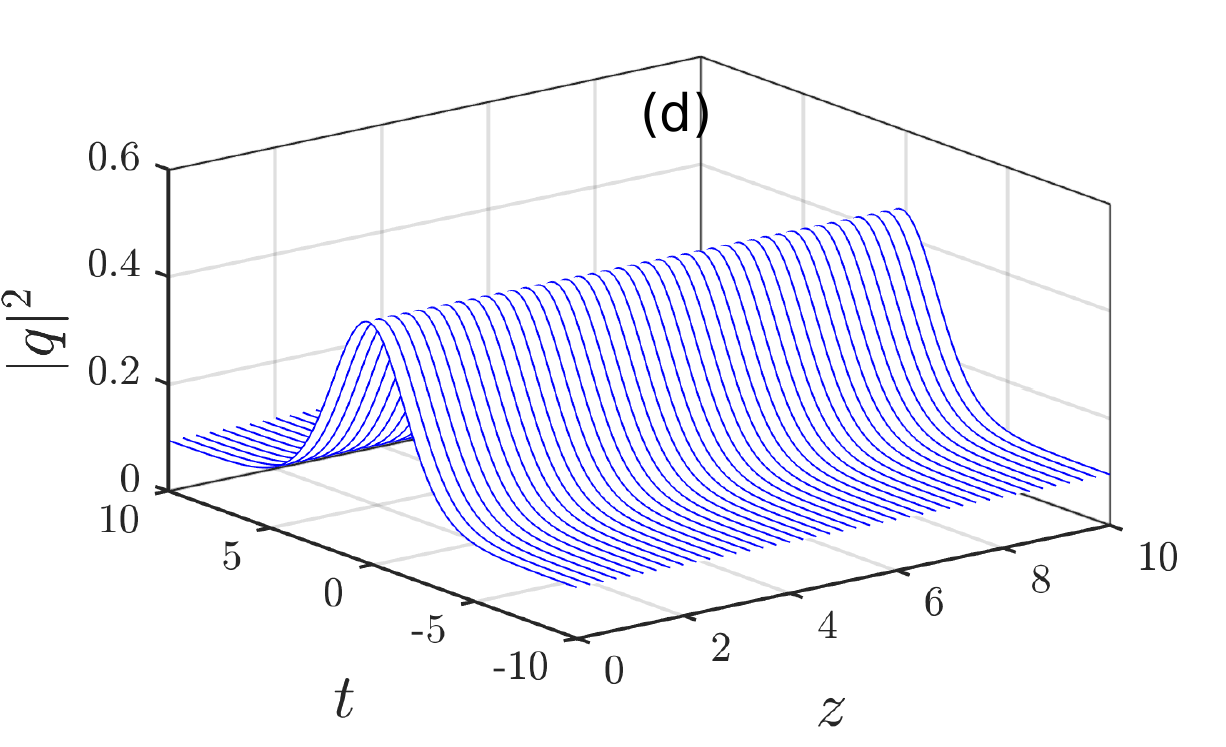}
	\caption{Plots show the intensity profiles of anti-dark solitary waves and their corresponding chirping profiles. The parameters are same as in Fig. 5 except $m=1$.}
\end{figure}
It is interesting to note that this constant value can be tuned to be zero by choosing the system parameters appropriately such that $\alpha^{2}=\frac{v_{2}}{v_{1}}$. This choice will render a bright solitary wave. Such a bright solitary wave for the choice of parameters $\kappa=0.02$, $k=0.322$, $u=0.01$ and $C=0.29$ is shown in Fig. 8. The stationary and propagation of bright solitary wave are portrayed in Figs. 8(a) and 8(c), respectively. Meanwhile, Fig. 8(b) depicts the  corresponding chirp profile. Even though the exact bright solitary wave has already been studied \cite{christ1} for the CQNLH system, the present form of bright solitary wave is unique in a mathematical sense. The reason is that the present bright solitary wave is obtained from the form of the shifted bright solitary wave by tuning solution parameters. As regards the corresponding chirping profile, as a striking feature, it exhibits a huge amplification near the leading and trailing edges of the chirping profile along with a well expanded flat profile around its maximum intensity. The numerical corroboration of the intensity evolution is also presented in Fig. 8(d). As reported in previous section, the role nonparaxiality does the same dynamics on the intensity and chirping profiles, which is not shown here. An important point to notice here is by altering chirp one can switch from anti-dark solitary wave to bright solitary wave.
\begin{figure}[ht]
	\includegraphics[width=0.45\linewidth]{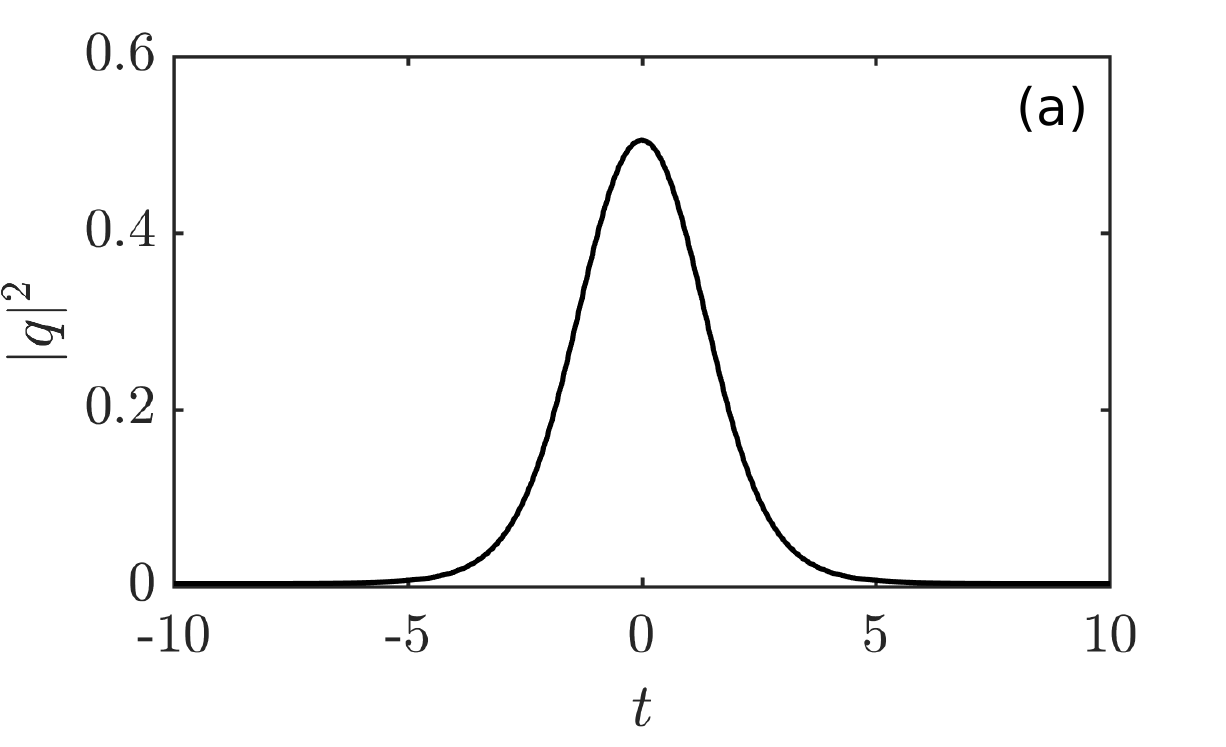}
	\includegraphics[width=0.45\linewidth]{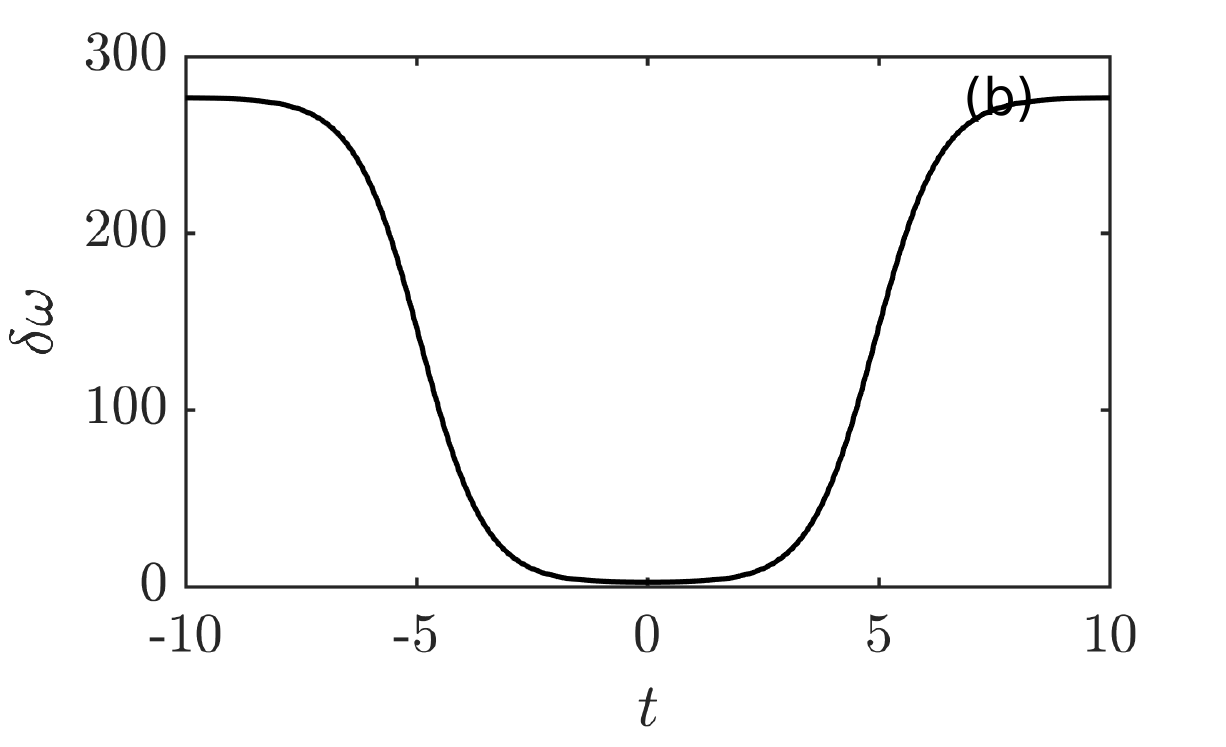}
	\includegraphics[width=0.45\linewidth]{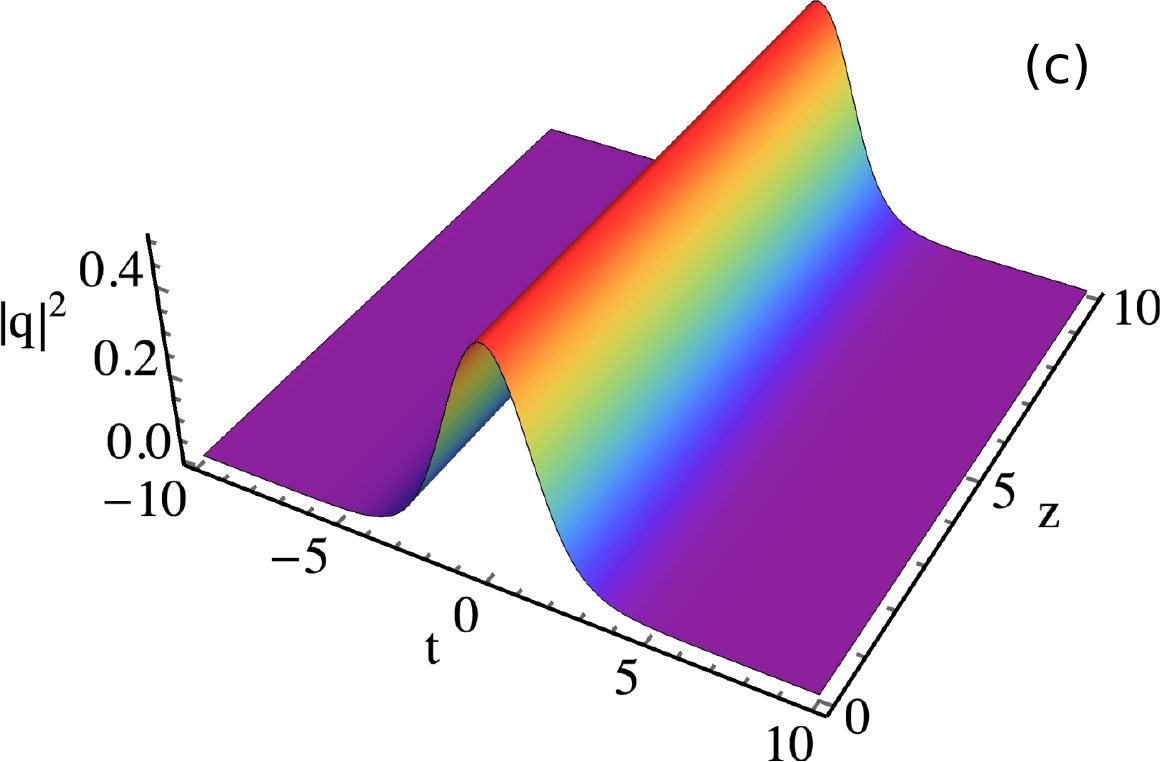}
	\includegraphics[width=0.5\linewidth]{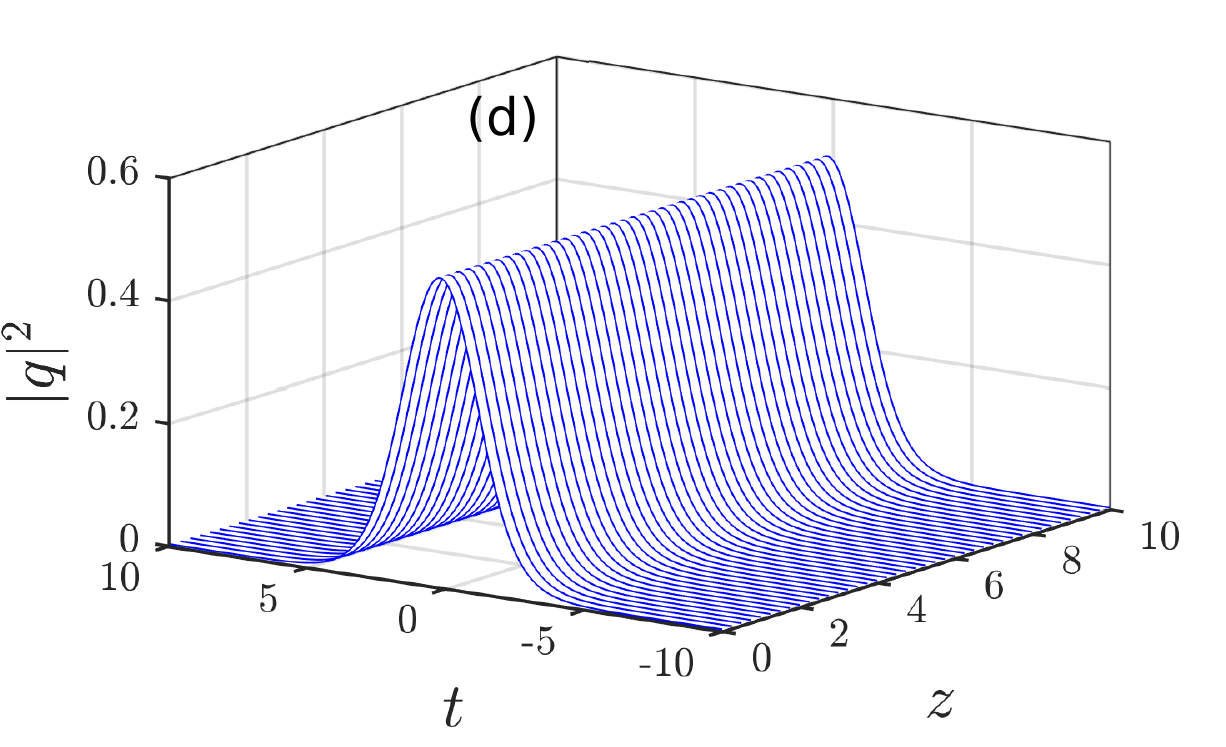}
	\caption{The propagation of bright solitary waves and their corresponding chirping profiles. Here too, the parameters are same as in Fig. 6 except $m=1$ and  $k=0.322$.}
\end{figure}
For completeness, below we present the periodic wave solution of Eq. \eqref{CQNLH} for the choice, $\alpha<0$ and $\beta>0$:
\bea\label{solution3}
q=\sqrt{\frac{v_{4}+v_{1} \tilde{\alpha}^{2}\sn^{2}(\tilde{u},m)}{1+\tilde{\alpha}^{2}\sn^{2}(\tilde{u},m)}}~e^{i(\phi(\xi)-k z)}, 
\eea
where $\tilde{u}=\sqrt{|d^{'}| (v_{4}-v_{2})(v_{4}-v_{3})}(\xi-\xi_{0})$ and the modulus parameter is $m=\frac{v_{1}-v_{2}}{v_{2}-v_{4}}\tilde{\alpha}^{2}$ (where $\tilde{\alpha}^{2}=\frac{v_{2}-v_{4}}{v_{1}-v_{3}}$). Note that this solution \eqref{solution3} reduces to the solitary waves solution of the CQNLS system given in Ref. \cite{crosta}, for the choice $\kappa\rightarrow 0$ and $m=1$. Apart from these solutions special bubble solutions of one-dimensional and higher dimensional CQNLS systems with competing nonlinearities have been reported in Refs. \cite{barashenkov,barashenkov1} but found to be unstable.
\subsection{Periodic wave with defocusing nonlinearities $(\alpha<0,\beta<0)$}
Next, we move on to construct the periodic waves of the CQNLH sysem \eqref{CQNLH} with defocusing nonlinearities i.e., $\alpha<0$ and  $\beta<0$. In order to construct the periodic wave solution, we adopt the same mathematical procedure as in the preceding section for Eq. \eqref{ode} (details are presented in Appendix B). The obtained periodic wave can be cast as
\bea\label{solution2}
q=\left(\frac{(v_{3}-v_{4} \tilde{\alpha}^{2} \sn^{2}(\tilde{u},m))}{1-\tilde{\alpha}^{2}\sn^{2}(\tilde{u},m)}\right)^\frac{1}{2}~e^{i(\phi(\xi)- k z)}.\,\,\,\,\,
\eea
where the argument of periodic wave is denoted as $\tilde{u}=\sqrt{|d^{'}| (v_{3}-v_{1}) (v_{3}-v_{2})}(\xi-\xi_{0})$ and the modulus parameter of the elliptic function $\sn$ is expressed as $m=(v_{1}-v_{4})/(v_{1}-v_{3}) \tilde{\alpha}^{2}$ (here $ \tilde{\alpha}^{2}=(v_{2}-v_{3})/(v_{2}-v_{4})$). Here too, one can notice that the solution is nonsingular as in the case of solution \eqref{solution1}, for all choice of system parameters satisfying the condition $v_{1}\geq v_{2}>v_{3}>v_{4}$. Thus, the above periodic wave \eqref{solution2} is characterized by four parameters. 
The amplitude is fixed by the four real roots. One can express the corresponding chirping parameter using Eq. \eqref{chirp} as
\bea\label{chirp02}
\delta\omega=-\left(\frac{u~(1-2~\kappa~k)}{s+2~u^{2}~\kappa}+\frac{c_{1}(1-\tilde{\alpha}^{2}\sn^{2}(\tilde{u},m))}{\left(v_{3}-v_{4} \tilde{\alpha}^{2} \sn^{2}(\tilde{u},m)\right)}\right).\,\,\,\,\,
\eea
Figures 9(a) and 9(c), respectively, denote the stationary and evolution of the periodic wave given by Eq. \eqref{solution2}. The pertinent chirping profile represented by Eq. \eqref{chirp02} is depicted in Fig. 9(b), whereas the numerical evolution of the periodic wave is shown in Fig. 9(d). The variation of intensity and chirp against the nonparaxial parameter is also shown in Fig. 10, where once again we observe that the chirp (intensity) decreases (increases) as the nonparaxial parameter increases.
\begin{figure}[t]
	\includegraphics[width=0.45\linewidth]{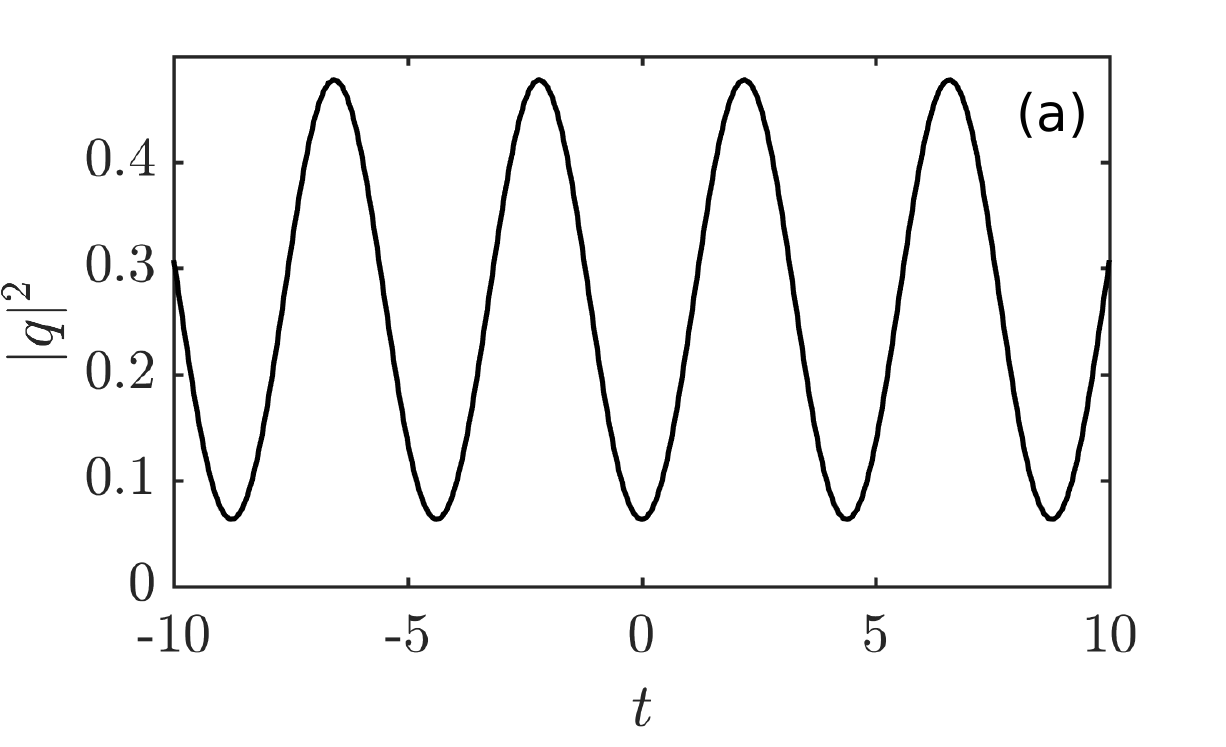}
	\includegraphics[width=0.45\linewidth]{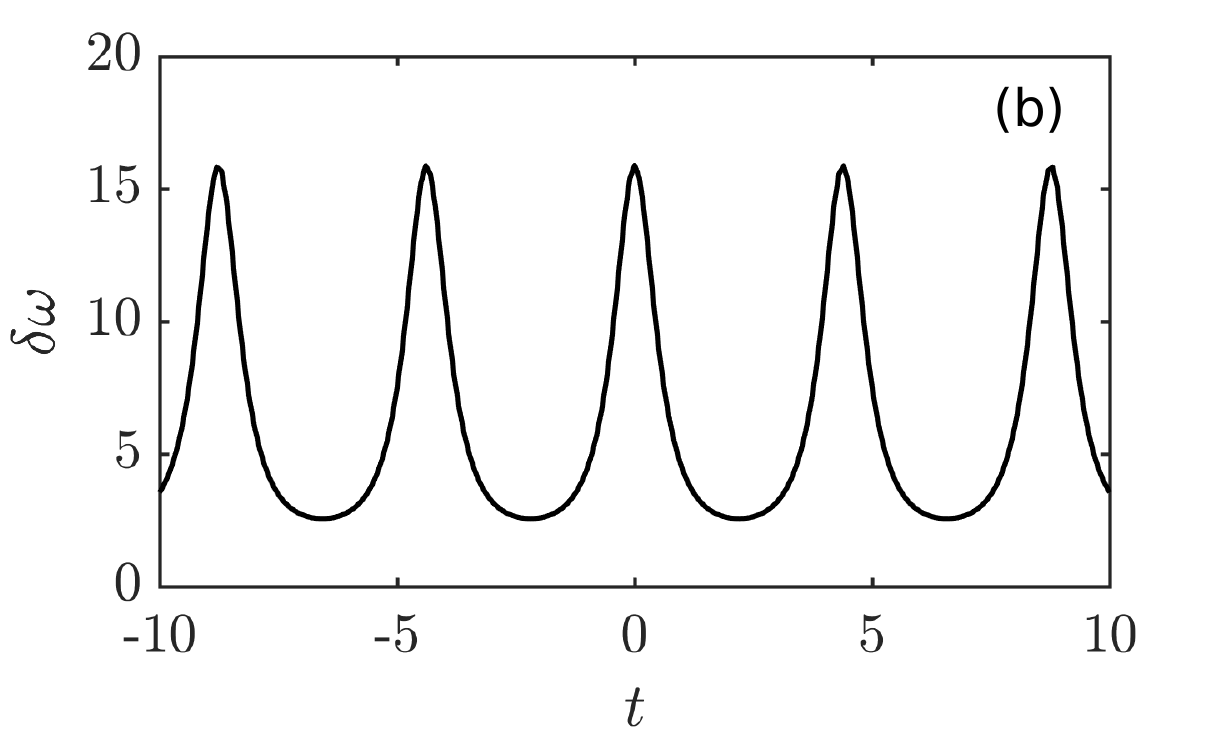}
	\includegraphics[width=0.45\linewidth]{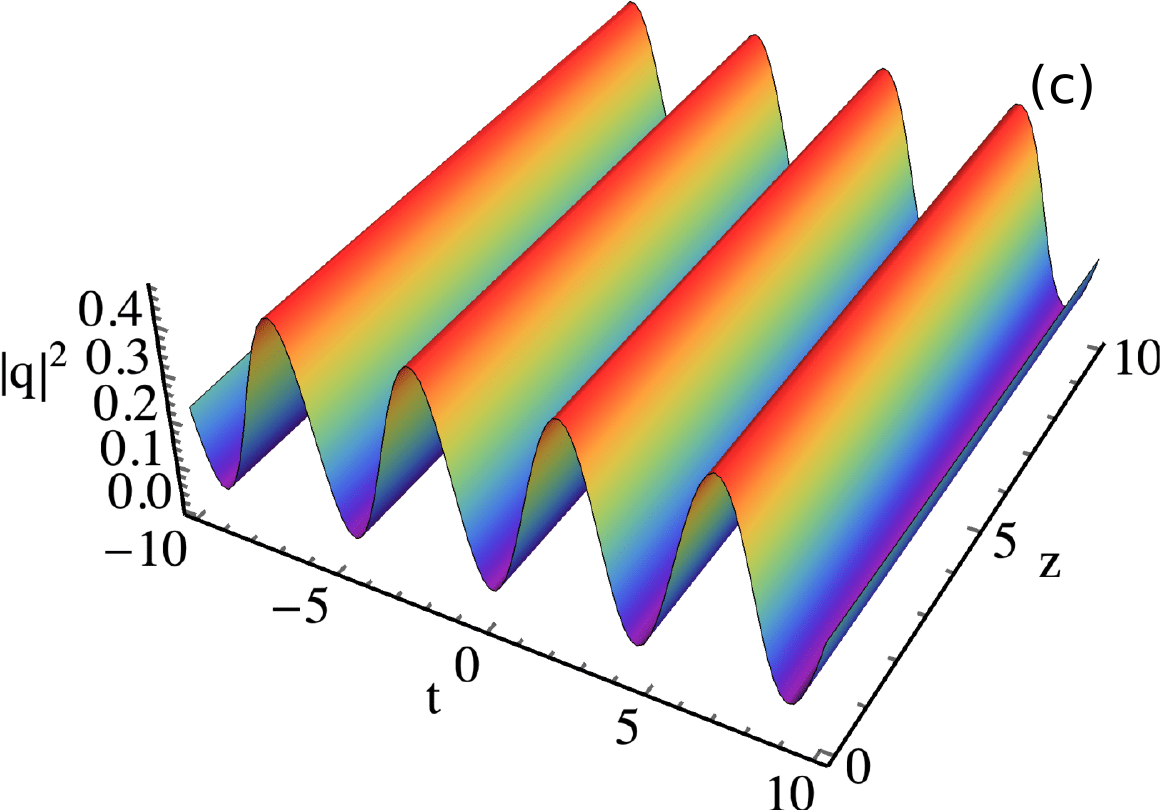}
		\includegraphics[width=0.5\linewidth]{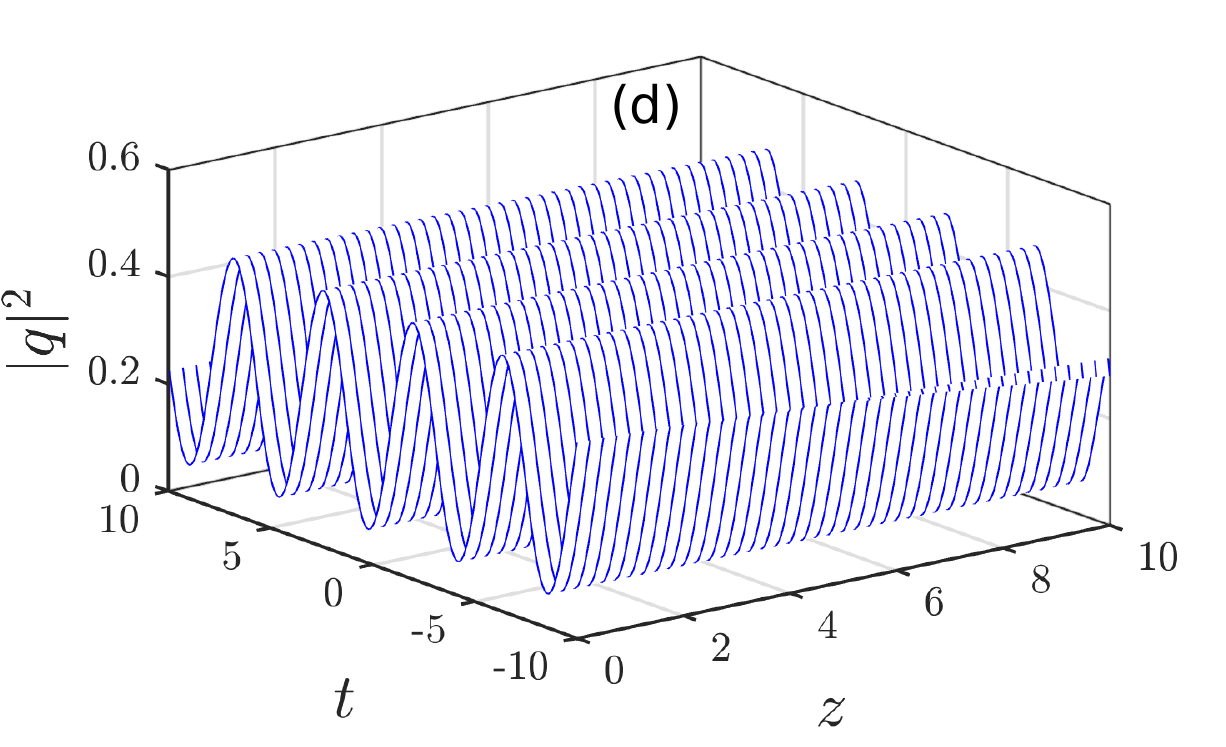}
	\caption{Plots depicting periodic waves \eqref{solution2} and their corresponding chirping profiles. The parameters are assigned as $\alpha=\beta=-1$, $\kappa=0.02$, $k=0.26$, $u=0.01$, $s=1$, $C=-0.25$, $c_{1}=1$ and $\xi_{0}=0$.}
	\end{figure}

\begin{figure}[t]
\includegraphics[width=0.55\linewidth]{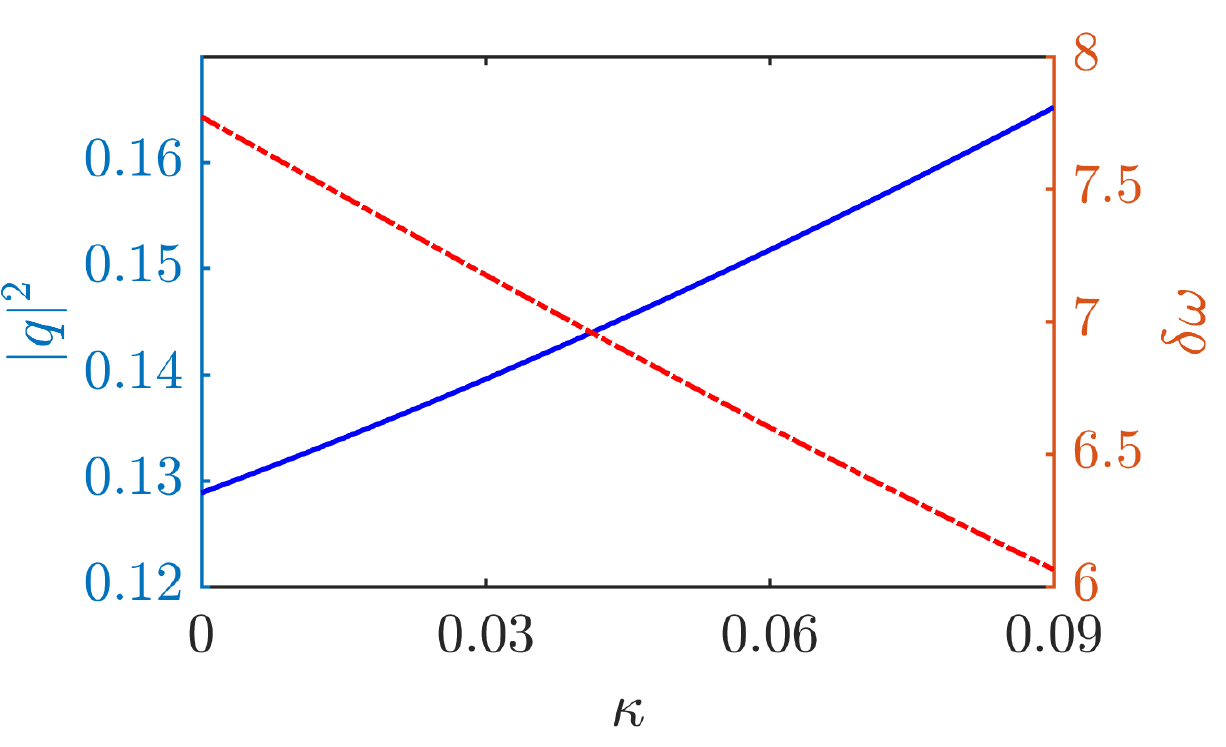}
\caption{The role of nonparaxial parameter on intensity and chirp profiles, which is obtained when the parameters are same as given in Fig. 9.}
\end{figure}
\subsection{Gray and Dark solitary waves ($\alpha<0$ and $\beta<0$)}
Finally, we study the gray and dark solitary wave solutions of the CQNLH system \eqref{CQNLH} for $\alpha<0$ and $\beta<0$. The periodic wave noted by Eq. \eqref{solution2} results in two types of solitary waves in the hyperbolic limit ($m=1$) with specific choice of system parameters. One can now at arrive the following solution by considering the condition $v_{1}=v_{2}$, $v_{3}>v_{4}$ \cite{byrd}
\bea\label{defocusing1}
q=\left(\frac{(v_{3}-v_{4} \tilde{\alpha}^{2} \tanh^{2}(\tilde{u}))}{1-\tilde{\alpha}^{2}\tanh^{2}(\tilde{u})}\right)^\frac{1}{2}~e^{i(\phi(\xi)- k z)}.\,\,\,\,\,
\eea
where $\tilde{u}=\sqrt{|d^{'}|} (v_{3}-v_{1}) (\xi-\xi_{0})$ with $m=1$. Here the inverse width of the solitary wave is given by $\sqrt{|d^{'}|} (v_{3}-v_{1})$ and the minimum dip of gray solitary wave is given by $v_{3}$. Note that $v_{1},v_{3},$ and $v_{4}$ are determined in terms of system parameters. One can clearly observe that the solitary wave solution \eqref{defocusing1} has nonzero asymptotic value given by $(\frac{v_{3}-v_{4} \alpha^{2}}{1-\alpha^{2}})$ when the variable $\xi$ approaches infinity ($|\xi| \rightarrow \infty$). The chirping parameter obtained through  Eq. \eqref{chirp} is given below
\bea
\delta\omega=-\left(\frac{u~(1-2~\kappa~k)}{s+2~u^{2}~\kappa}+\frac{c_{1}(1-\tilde{\alpha}^{2}\tanh^{2}(\tilde{u}))}{\left(v_{3}-v_{4} \tilde{\alpha}^{2} \tanh^{2}(\tilde{u})\right)}\right).\,\,\,\,\,
\eea
The propagation of gray solitary wave is shown in Fig. 11(c) (which is further confirmed numerically in Fig. 11(d)) while the stationary solution is portrayed in Fig. 11(a) (the minimum intensity does not drop to zero at the dip center). Here too, the solution \eqref{solution2} has the negative chirping frequency as shown in Figs. 11(b).

We note that for $v_{3}=0$, the minimum dip drops down to zero thereby resulting in dark solitary wave. Such a dark solitary wave is dipicted in Fig. 12 for the choice of parameters $\kappa=0.02$, $k=0.25$, $u=0.01$, $s=1$, $C=-0.378$ and $c_{1}=1$.  Figures 12(a) and 12(c), respectively, show the stationary and propagating dark solitary waves in the CQNLH system and the numerical simulation showing the evolution of dark solitary wave is presented in Fig. 12(d). Indeed, we have verified that the gray solitary wave remains stable for longer propagation distance at $z=20$. In waveguide geometry $z=10$ is sufficient to perform experiments. A remarkable outcome that one can observe is the unusual chirping profile (see Fig. 12(b)) pertaining to the dark solitary wave, wherein the chirped profile undergoes a significant compression accompanied by a huge amplification in its amplitude which is not reported anywhere in the literature for the NLH system.
\begin{figure}[ht]
\includegraphics[width=0.45\linewidth]{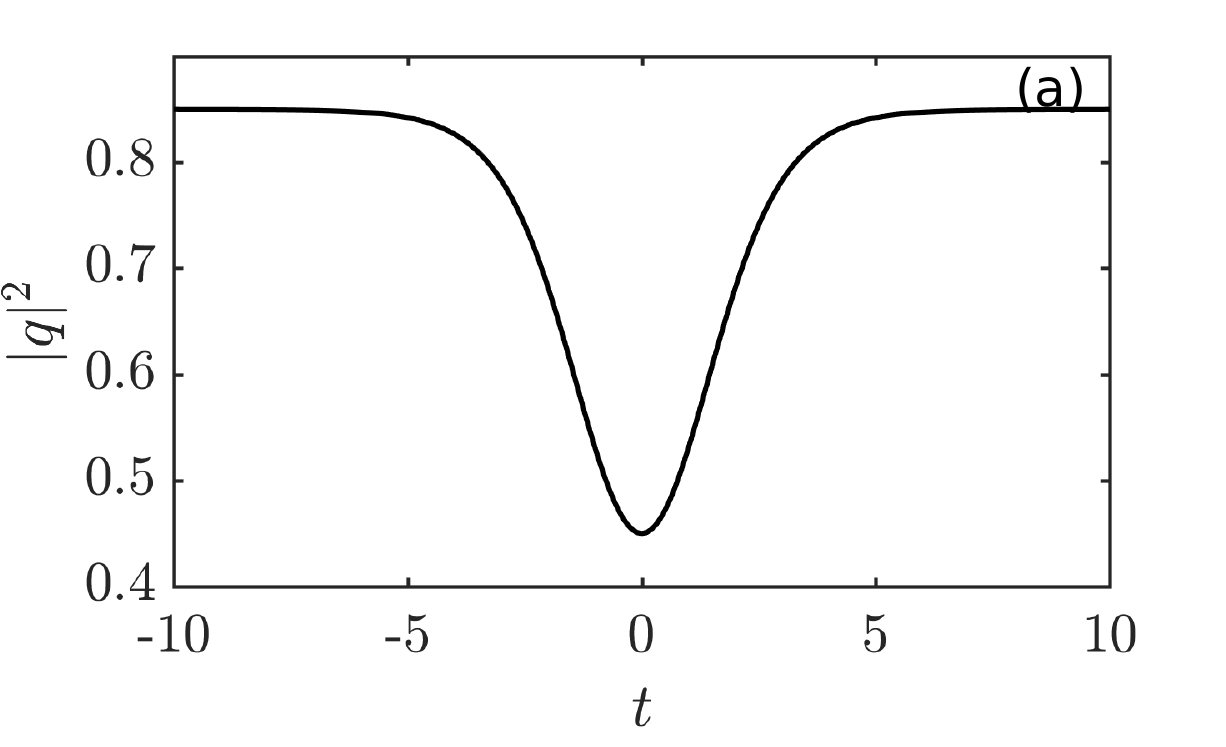}
\includegraphics[width=0.45\linewidth]{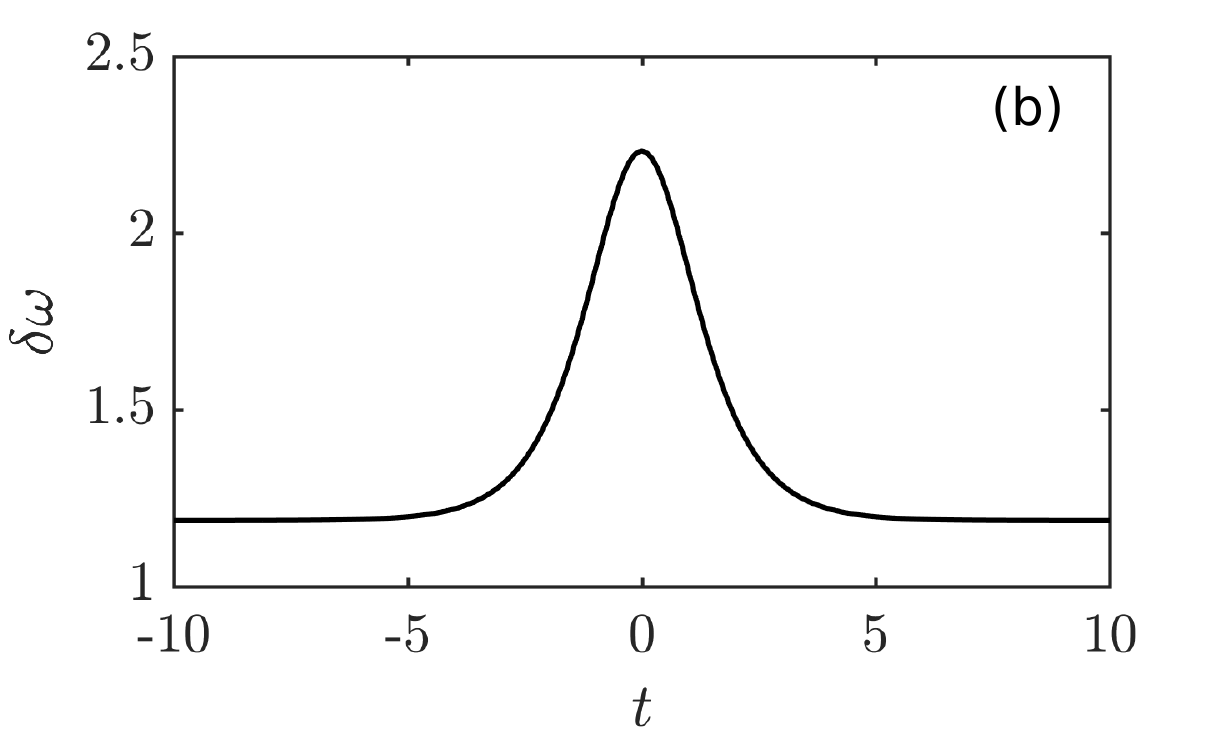}
\includegraphics[width=0.45\linewidth]{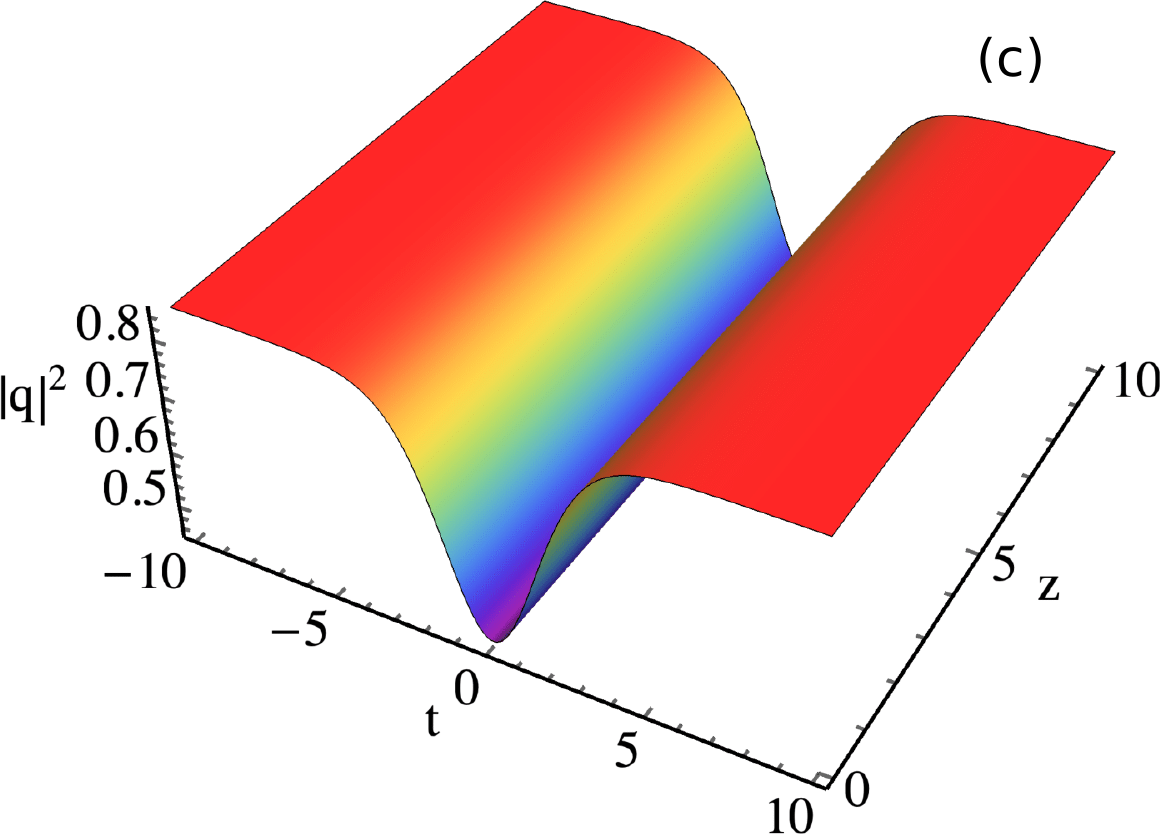}
\includegraphics[width=0.5\linewidth]{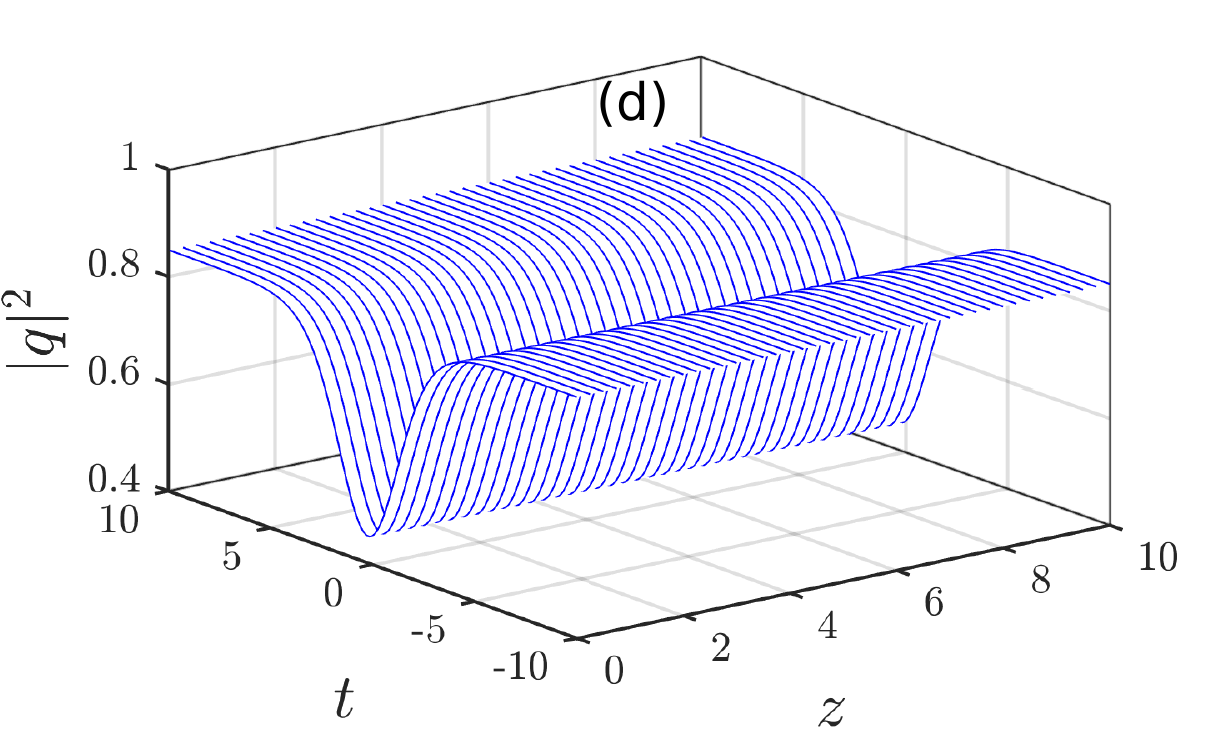}
\caption{The evolution of gray solitary waves and their corresponding chirping profiles. The parameters are same as given in Fig.~10 except $k=0.15$, $C=-0.1$ and $m=1$.}
\end{figure}
\begin{figure}[ht]
\includegraphics[width=0.45\linewidth]{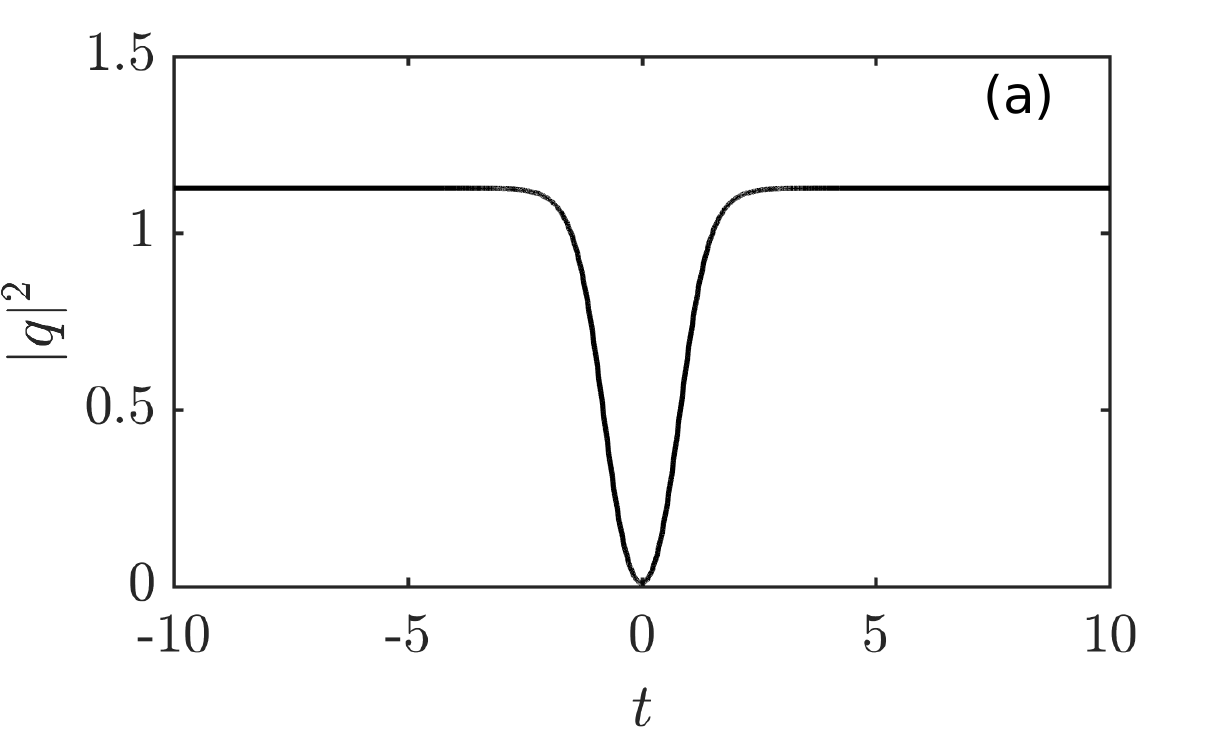}
\includegraphics[width=0.45\linewidth]{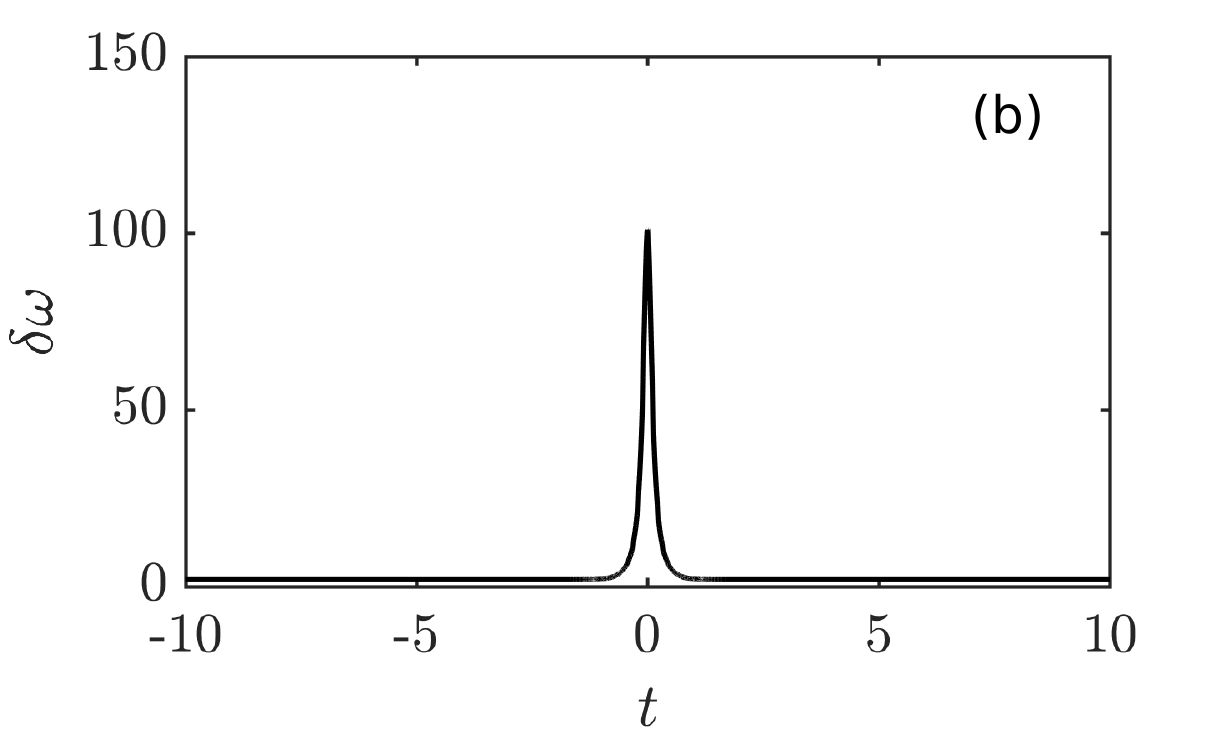}
\includegraphics[width=0.45\linewidth]{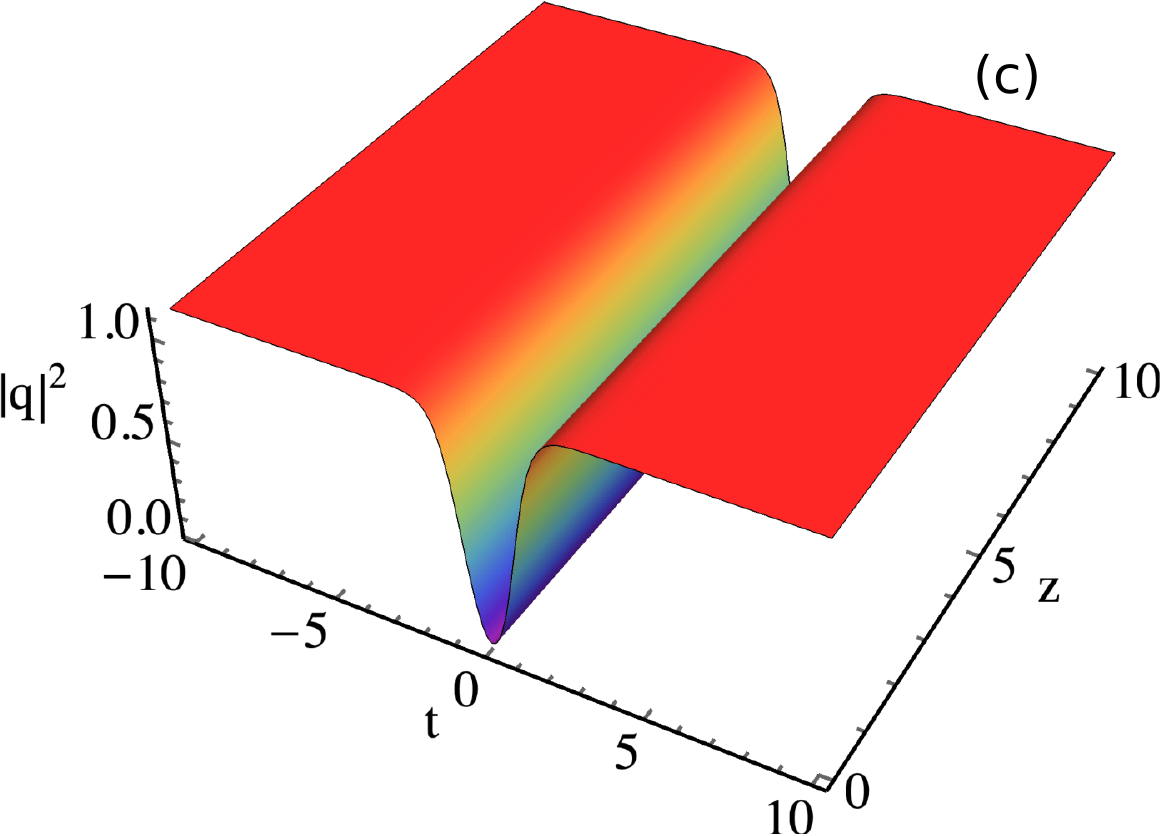}
\includegraphics[width=0.5\linewidth]{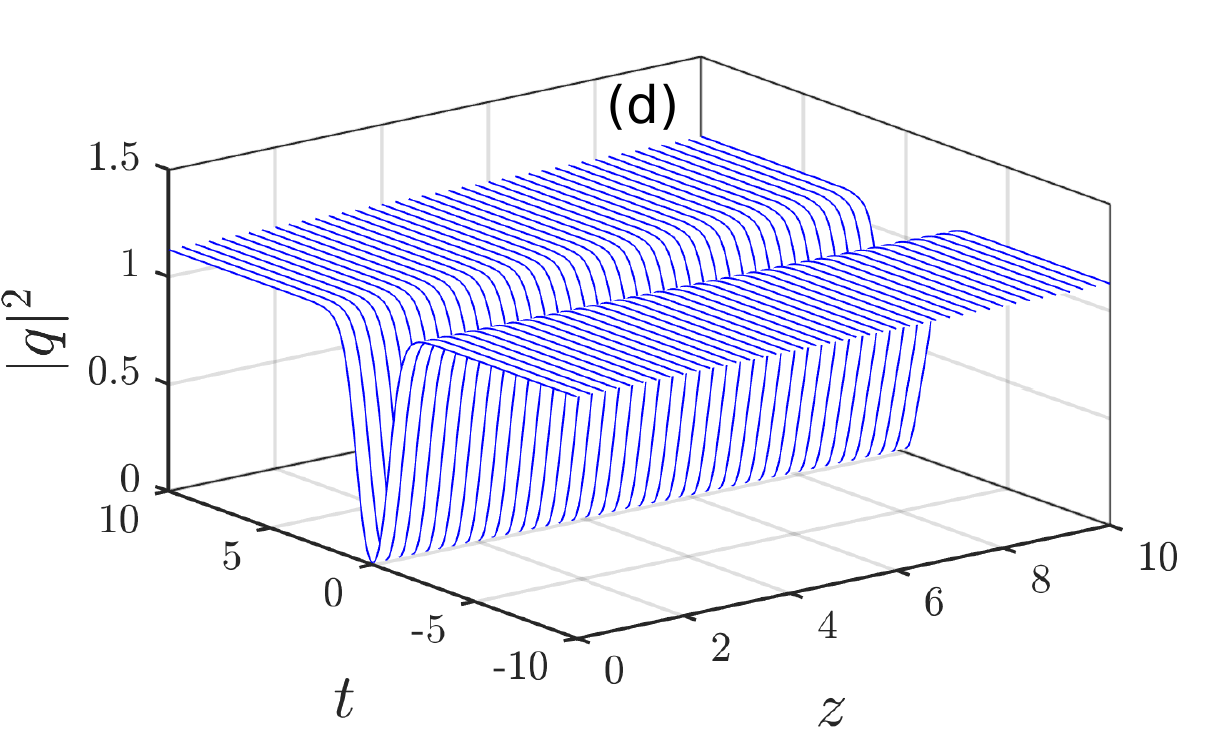}
	\caption{Intensity profiles of dark solitary waves and their corresponding chirping profiles. The parameters are same as in Fig.~10 except $k=0.25$, $C=-0.378$ and $m=1$.}
\end{figure}

\section{conclusion}
\label{sec5}
We have considered a scalar nonlinear envelope equation with spatio-temporal dispersion and cubic-quintic nonlinearities. In the first part, we have performed the MI analysis of a plane wave for the CQNLH system by using a linear stability analysis. The influence of various physical parameters such as power, nonlinearities (both cubic and quintic) and the nonparaxial parameter on the MI gain spectra have been discussed in detail. In particular, the study unraveled an interesting feature of the nonparaxiality, which lead to a  suppression in the conventional gain spectrum as the value of nonparaxiality increases. Also, a nontrivial monotonically increase in gain spectrum formed near the tails of the conventional MI band. Besides, we have shown the generation of train of ultra-short nonparaxial pulses in order to emphasize the long time behaviour of the perturbed CW solution by employing systematic numerical simulations.

Secondly, we have obtained a family of periodic waves for distinct physical settings in the CQNLH system including competing ($\alpha > 0, \beta < 0$) and defocusing nonlinearities ($\alpha < 0, \beta < 0$). In the case of hyperbolic limit, this set of periodic waves turns out to be the physically interesting chirped anti-dark, bright and gray, dark solitary waves. In particular, for the competing nonlinearities the anti-dark solitary results for the chioce $v_{3}=v_{4}$, $v_{1}>v_{2}$ which results in bright solitary wave when $\tilde{\alpha}^{2}=(\frac{v_{2}}{v_{1}})$. Then for the defocusing nonlinearities the choice $v_{1}=v_{2}$, with $v_{3}>v_{4}$ lead to gray solitary wave that turns out to be dark solitary wave for $v_{3}=0$. The nontrivial phase chirping of these nonlinear waves that varies inversely as a function of intensity is a distinct feature of this CQNLH system. Specifically, we have reported two unusual chirping profiles related to bright and dark solitary waves experiencing some unusual nonlinear dynamics including compression and amplification, a novel outcome which is not observed in any NLH system. Such a huge chirping even with small nonparaxial parameter $\kappa$ is a striking feature of CQNLH system that will find applications in pulse/beam shaping. Also, we have shown the intensity (chirp) of the periodic waves gets decreased (increased) with the increase of the nonparaxial parameter $\kappa$. All of our analytical solutions are well confirmed by direct numerical simulations. We hope that these ramifications may find applications in miniaturized photonic devices, telecommunications  fibers and plasmonics. We note here that in a general perspective, there arise several future directions in view of the preset work, for instance, it is of natural interest to investigate the CQNLH system in the presence of higher order linear and nonlinear effects including self-steepening, Raman frequency shift and third order dispersion. Then attempts to modify the CQNLH system in the presence of the GRIN medium involving spatially or spatio-temporally modulated nonlinearity and looking for the possibility of chirped solitons could be of considerable research interest. Apart from these, one can also consider the chirped nonparaxial solitons and rogue waves in the multi-component CQNLH system and their generalizations in higher dimensions. Works are in progress along these directions.

\section*{Acknowledgement}
The work of K.T. is supported by a Senior Research Fellowship from Rajiv Gandhi National Fellowship for SC candidates, University Grants Commission (UGC), Government of India. The work of T.K. is supported by Department of
Science and Technology-Science and Engineering Research Board (DST-SERB), Government
of India, in the form of a major research project (File No. EMR/2015/001408). A.G. wishes to thank DST-SERB,  Government of India,  for providing a National Postdoctoral Fellowship (Grant No. PDF/2016/002933).
\appendix
\section{Procedure for solving the CQNLH system for competing nonlinearities ($\alpha>0$ and $\beta<0$)}
We start by rewriting Eq. \eqref{ode} with $\alpha>0$ and $\beta<0$
\bea\label{ode01}
\frac{1}{4}\left(\frac{d(\psi)^{2}}{d\xi}\right)^{2}=a^{'} \psi^{2}-b^{'}~\psi^{4}-c^{'}~\psi^{6}+d^{'}\psi^{8}+a_{0}^{'},\nonumber\\
\eea
where $a^{'}=2C$, $C$ being an integration constant, $a_{0}^{'}=c_{1}^{2}$, $b^{'}=\left(b/a\right)$, $c^{'}=\left(|\alpha|/a\right)$ and $d^{'}=\left(|\beta|/a\right)$. We arrive at the following expression by integrating Eq.~(\ref{ode01})
\bea\label{ode1}
2\sqrt{d^{'}} (\xi-\xi_{0})=\int \frac{dv}{\sqrt{(v-v_{1})(v-v_{2})(v-v_{3})(v-v_{4})}},\nonumber\\
\eea
where $v=\psi^{2}$, $\xi_{0}$ is an initial value of co-ordinate $\xi$  and $v_{1}$, $v_{2}$, $v_{3}$ and $v_{4}$ are the four roots of the following quartic equation
\bea\label{ode2}
v^{4}-\frac{c^{'}}{d^{'}}v^{3}-\frac{b^{'}}{d^{'}}v^{2}+\frac{a^{'}}{d^{'}}v+\frac{a_{0}^{'}}{d^{'}}=0.
\eea
In order to solve Eq. \eqref{ode2}, we first perform the transformation, $v=x+c^{'}/4 d^{'}$ and after regrouping the terms, we obtain the depressed quartic equation as
\bea\label{ode3}
x^{4}+ p x^{2}+q x +r =0,
\eea
where,
\bes\bea
p=-\frac{3c^{'2}}{8d^{'2}}-\frac{b^{'}}{d^{'}},\\
q=-\frac{c{'3}}{8d^{'3}}-\frac{c^{'}b^{'}}{2d^{'2}}+\frac{a^{'}}{d^{'}},\\
r=\frac{-3c^{'4}}{256d^{'4}}-\frac{b^{'}c^{'2}}{16 d^{'3}}+\frac{a^{'}c^{'}}{4d^{'2}}+\frac{a_{0}^{'}}{d^{'}}.
\eea\ees
One can easily obtain the following roots by employing straightforward calculation of the depressed quartic equation \eqref{ode3},
\bea\label{roots}
x_{1-4}=\frac{\pm_{1}\sqrt{2y}\pm_{2}\sqrt{-(2(p+y)\pm_{1}\frac{\sqrt{2}q}{\sqrt{y}})}}{2}.
\eea
To find the roots of Eq. \eqref{ode3}, we fix the discriminant $q^{2}-2\sqrt{(y^{2}+y p+p^{2}/4-r)(2y)}$ to be zero, where $y$ is a root of the equation. 
  Now one can write the four roots of the quartic equation \eqref{ode2} as $v_{1-4}=\left(x_{1-4}+c^{'}/4 d^{'}\right)$. Next we assume that $v$ to be real in Eq. \eqref{ode2} and choose its four distinct roots as $v_{1}>v_{2}>v_{3}\geq v_{4}$. We restrict the case for which at least three roots are positive $v_{1}> v_{2}>v_{3}>0$ and these roots lie between $v_{2}<v<v_{1}$. To solve the expression \eqref{ode1}, we introduce a new variable $\tilde{t}$ instead of $v$ in which  $\tilde{t}=\sqrt{(v_{2}-v)/(v_{1}-v)}$ and the expression (\ref{ode1}) becomes the first kind of elliptic integral as \cite{byrd}
\bea\label{integral}
\tilde{u}=\int_0^{\tilde{t}}\frac{d\tilde{t}}{\sqrt{(1-\theta^{2})(1-m \theta^{2})}}
\eea
where
$\theta=\tilde{t}/\tilde{\alpha}$, the modulus parameter denoted as $m=\tilde{\alpha}^{2}(v_{1}-v_{4})/(v_{2}-v_{4})$ $(0\leq m \leq 1$), $\tilde{\alpha}^{2}=(v_{2}-v_{3})/(v_{1}-v_{3})$, and $\tilde{u}=\sqrt{|d^{'}| (v_{2}-v_{3}) (v_{2}-v_{4})}(\xi-\xi_{0})$. Finally, we can express the solution of standard elliptic integral Eq. \eqref{integral} in terms of Jacobi elliptic function as follows.
\bea\label{sol_v}
\psi=\sqrt{\frac{v_{2}-v_{1} \alpha^{2}\sn^{2}(\tilde{u},m)}{1-\alpha^{2}\sn^{2}(\tilde{u},m)}}.
\eea
Then from Eq. \eqref{ansatz1}, the solution of Eq. \eqref{CQNLH} with competing nonlinearities is given by
\bea
q=\left(\frac{(v_{2}-v_{1} \tilde{\alpha}^{2} \sn^{2}(\tilde{u},m))}{1-\tilde{\alpha}^{2}\sn^{2}(\tilde{u},m)}\right)^\frac{1}{2}~e^{i(\phi(\xi)- k z)}.
\eea
The analysis of this solution is carried out in Sec. IVA.
\section{Procedure for solving the CQNLH with $\alpha<0$ and $\beta<0$}
Here, we present the procedure to construct periodic wave solution for the CQNLH system \eqref{CQNLH} with defocusing nonlinearities, i.e., ($\alpha<0$ and $\beta<0$). In this case Eq. \eqref{ode} becomes as
\bea\label{ode4}
\frac{1}{4}\left(\frac{d(\psi)^{2}}{d\xi}\right)^{2}=a^{'} \psi^{2}-b^{'}~\psi^{4}+c^{'}~\psi^{6}-d^{'}\psi^{8}+a_{0}^{'}.\nonumber\\
\eea
Following the mathematical procedure as given in the preceding section (Appendix A), we solve Eq. \eqref{ode4}. Then the solution can be expressed in terms of Jacobi elliptic function as follows.
\bea\label{sol_v1}
\psi=\sqrt{\frac{v_{3}-v_{4} \alpha^{2}\sn^{2}(\tilde{u},m)}{1-\alpha^{2}\sn^{2}(\tilde{u},m)}}.
\eea
During the construction of this solution \eqref{sol_v1}, we assume the four distinct roots as $v_{1}\geq v_{2}>v_{3}> v_{4}$ and restrict at least two roots are to be positive $(v_{2}> v_{3}>0)$. Hence the roots to lie between $v\in[v_{2},v_{3}]$. Finally, one can express the periodic wave for this defocusing CQNLH system as
\bea
q=\left(\frac{(v_{3}-v_{4} \tilde{\alpha}^{2} \sn^{2}(\tilde{u},m))}{1-\tilde{\alpha}^{2}\sn^{2}(\tilde{u},m)}\right)^\frac{1}{2}~e^{i(\phi(\xi)- k z)}.
\eea
This solution is also discussed in detail in Sec. IVC.

\end{document}